\documentclass[10pt, conference, compsocconf]{IEEEtran}
\usepackage{cite}

\ifCLASSINFOpdf
\else
\fi
\usepackage{algorithm}
\usepackage[noend]{algpseudocode}
\usepackage{amsfonts}
\usepackage{amsthm}
\usepackage{amsmath}
\usepackage{wrapfig}
\usepackage{graphicx}
\usepackage{subcaption}
\usepackage{multirow}
\usepackage{bm}
\usepackage[draft]{todonotes}
\usepackage{float}
\DeclareMathOperator{\diag}{diag}
\DeclareMathOperator*{\argmin}{arg\,min}

\newcommand*{\data}{./data}

\usepackage{mathtools}
\DeclarePairedDelimiter{\ceil}{\lceil}{\rceil}

\begin{document}
%
\title{Avoiding Synchronization in First-Order Methods for  Sparse Convex Optimization}


\author{\IEEEauthorblockN{Aditya Devarakonda}
\IEEEauthorblockA{EECS Department\\
Univeristy of California, Berkeley\\
aditya@eecs.berkeley.edu}\\
\IEEEauthorblockN{Kimon Fountoulakis}
\IEEEauthorblockA{ICSI and Statistics Department\\
Univeristy of California, Berkeley\\
kfount@stat.berkeley.edu}
\and
\IEEEauthorblockN{James Demmel}
\IEEEauthorblockA{EECS and Math Department\\
Univeristy of California, Berkeley\\
demmel@eecs.berkeley.edu}\\
\IEEEauthorblockN{Michael W. Mahoney}
\IEEEauthorblockA{ICSI and Statistics Department\\
Univeristy of California, Berkeley\\
mmahoney@stat.berkeley.edu}
}


%


\maketitle

\begin{abstract}
Parallel computing has played an important role in speeding up convex optimization methods for big data analytics and large-scale machine learning (ML). However, the scalability of these optimization methods is inhibited by the cost of communicating and synchronizing processors in a parallel setting. Iterative ML methods are particularly sensitive to communication cost since they often require communication every iteration. In this work, we extend well-known techniques from Communication-Avoiding Krylov subspace methods to first-order, block coordinate descent methods for Support Vector Machines and Proximal Least-Squares problems. Our Synchronization-Avoiding (SA) variants reduce the latency cost by a tunable factor of $s$ at the expense of a factor of $s$ increase in flops and bandwidth costs. We show that the SA-variants are numerically stable and can attain large speedups of up to $5.1\times$ on a Cray XC30 supercomputer.
\end{abstract}

\begin{IEEEkeywords}
Synchronization-Avoiding;
Support Vector Machines;
Proximal Least-Squares;
Sparse Convex Optimization;
Coordinate Descent Methods;
\end{IEEEkeywords}

%
\IEEEpeerreviewmaketitle


\section{Introduction}
\begin{figure*}[t!]
  \centering
  \includegraphics[width=.8\textwidth]{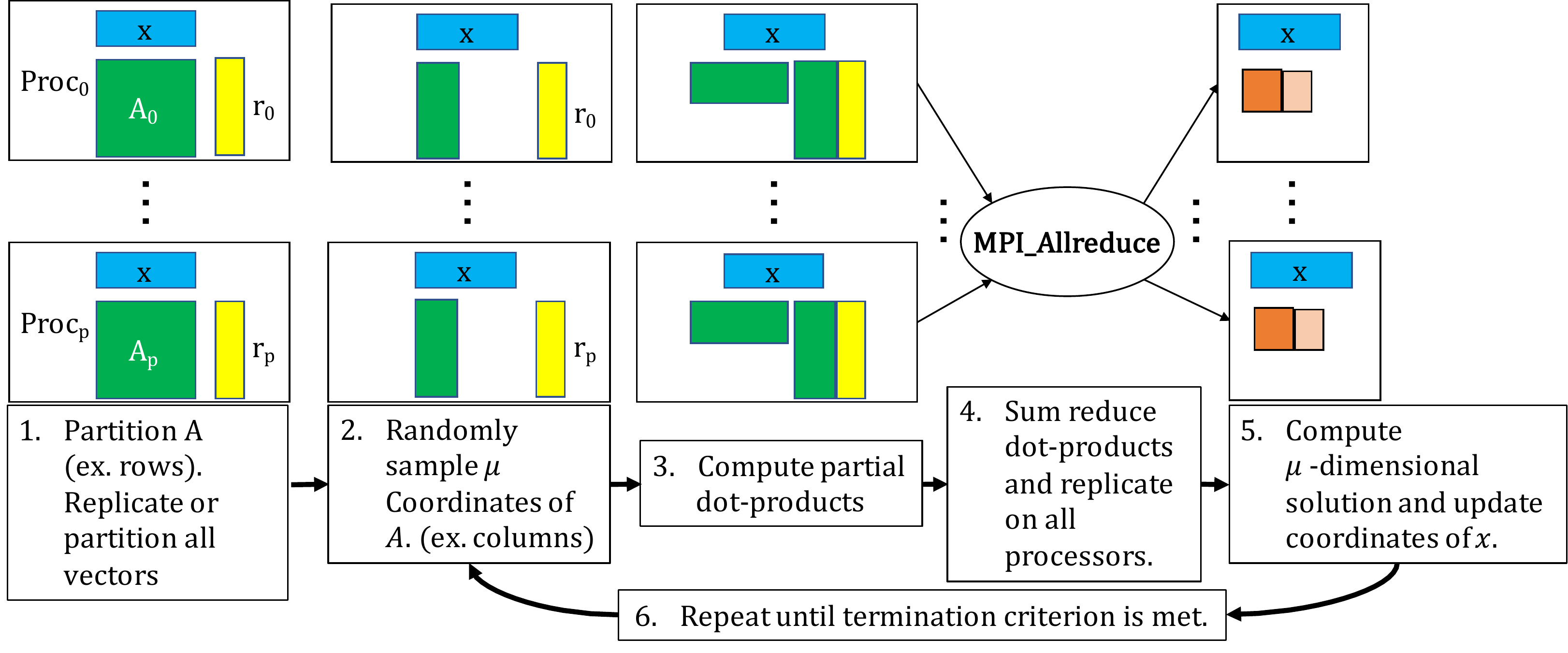}
  \caption{A high-level depiction of the Block Coordinate Descent method (independent of the minimization problem being solved). The matrix $A$ is $1D$-row partitioned and w.l.o.g. depicted as being dense. Vectors in the partitioned dimension are also partitioned (in this case residual vector, $r$). Vectors in the non-partitioned dimension and all scalars are replicated (in this case $x$). Each processor selects the same column indices (by using the same random generator seed). Computation of (partial) dot-products is a local GEMM operation. After that, the results are combined using an Allreduce with summation. Due to data replication all processors can independently compute this iteration's solution and perform vector updates.}
  \label{fig:cdag}
\end{figure*}
The running time of algorithms on a distributed-memory, parallel machine depends on the number of arithmetic operations $F$ (computation) and the cost of data movement (communication). 
In distributed-memory settings, communication cost includes the ``message size cost," i.e., the amount, $W$, of data exchanged between processors over a network (bandwidth cost) and the ``synchronization cost," i.e., the number, $L$, of messages sent, where a message is used for interprocessor synchronization (latency cost). 
On modern parallel computer architectures, communicating data often takes much longer than performing a floating-point operation, and this gap is continuing to increase. 
Therefore, it is important to design algorithms that minimize communication in order to attain high performance, and recent results on communication-avoiding numerical linear algebra (CA-NLA) \cite{ballard14} illustrate the potential for faster algorithms by avoiding interprocessor communication.

Our goal in this work is to extend results from CA-NLA to convex optimization \cite{boyd04} to accelerate machine learning (ML) applications. In particular, we are interested in reducing communication cost for iterative convex optimization on distributed machines. Communication is a bottleneck for these methods since they often require synchronization at every iteration. Our goal in this paper is to avoid this communication for $s$ iterations, where $s$ is a tuning parameter, without altering the convergence rates or numerical stability of the existing methods. We will show how the communication-avoiding techniques developed in \cite{ballard14} can be used to derive faster convex optimization algorithms.

We are interested in solving the class of optimization problems which take the form
\begin{align}
\argmin_{x \in \mathbb{R}^n} f(A, b, x) + g(x)\label{eq:opt}.
\end{align}
where $f(A, b, x)$ is a convex loss function with regularization function, $g(x)$, $A \in \mathbb{R}^{m \times n}$ with $m$ data points and $n$ features, labels $b \in \mathbb{R}^m$, and the solution to the minimization problem, $x \in \mathbb{R}^n$. In particular, we would like to focus on sparse convex optimization problems: sparse proximal least-squares \cite{parikh14}, $\left(f(A,b,x) = \frac{1}{2n} ||Ax - b ||_2^2\right)$ with sparsity-inducing regularization functions
\begin{align*}
&\text{Lasso \cite{tibshirani96}:}&g(x) &= \lambda||x||_1,\\
&\text{Elastic-Nets \cite{zou05}:}&g(x) &= \lambda||x||_2^2 + (1-\lambda)||x||_1,\\
&\text{Group Lasso \cite{yuan06}:}&g(x) &=  \lambda\sum\nolimits_{g = 1}^G ||\tilde x_g||_2,
\end{align*}
where $\{\tilde x_1, \tilde x_2, \ldots, \tilde x_G\}$ are $G$ disjoint blocks of $x$.

We also consider support vector machines (SVMs)  \cite{cortes95} 
$$f(A,b,x)  = \lambda \sum\nolimits_{i =1}^m \max(1 - b_iA_ix, 0)^2 , $$
where $A_i$ is the $i$-th row of $A$, $b_i$ is the corresponding binary label ($\{+1, -1\}$), and $\lambda$ is a regularization parameter. We present our results for proximal least-squares using Lasso-regularization, but they hold more generally for other regularization functions with well-defined proximal operators (Elastic-Nets, Group Lasso, etc.) \cite{parikh14}. By proximal operator we mean that the non-linearity due to the regularization function, for example with Lasso, can be defined for a constant, $\alpha$, and element-wise on vector, $\beta$, as:
\begin{align}
S_{\alpha}(\beta_i) := sign(\beta_i)\max(|\beta_i| - \alpha, 0), \label{eq:sth}
\end{align}
where $S_\alpha(\beta)$ is the well-know soft-thresholding operator for Lasso \cite{tibshirani96,beck09, daubechies04}. Proximal operators, similar to \eqref{eq:sth}, can be defined for other non-linear regularizers (i.e. Elastic-Nets, Group Lasso, etc.). Lasso creates solution sparsity during the optimization process by setting elements of the solution vector, $x$, exactly to zero. SVM \cite{cortes95} introduces sparsity (by its loss function definition) since we seek a small number of support vectors which separate data belonging to different classes. 
The sparsity-inducing nature of both optimization problems is important (and widely used) when dealing with high-dimensional data.
In the case of high-dimensional and large-scale data, it becomes equally important to parallelize the computations in order to quickly and efficiently solve these Lasso and SVM problems.

There exist many algorithms to solve proximal least-squares \cite{nesterov12, fercoq15, richtarik14, richtarik16, atsushi14} and SVM problems \cite{platt98, chang08, pegasos, sorsvm,dcdsvm08}. In this paper we focus on randomized variants of accelerated and non-accelerated Coordinate Descent (CD) and Block Coordinate Descent (BCD) since they have optimal convergence rates among the class of first-order methods \cite{fercoq15, nesterov12, richtarik14, richtarik16, dcdsvm08,chang08}.

CD \cite{Wright15} is a popular ML technique for solving optimization problems which updates a single element (i.e., coordinate) of $x$ by minimizing a one-dimensional subproblem. BCD generalizes CD with a tunable block size parameter, $\mu$, by updating $\mu$ coordinates of $x$ and minimizing a $\mu$-dimensional subproblem. This process is repeated until a termination criterion is met. Solving the subproblem initially requires the computation of several dot-products with a subset of rows or columns of the data matrix, $A$, its transpose, and residual vectors. Once they are computed the remaining computations are scalar and vector operations which update the solution and residual vectors (and scalar quantities) for the next iteration. 

Figure \ref{fig:cdag} illustrates the computations in parallel BCD. We assume that $A$ is partitioned so that the dot-products can be computed in parallel by all processors (in this case row-partitioned). Vectors in $\mathbb{R}^m$ are similarly partitioned (in this case residual, $r$), but vectors in $\mathbb{R}^n$ are replicated (in this case solution, $x$). For $\mu > 1$ the dot-products become GEMM (GEneralized Matrix-Multiplication) computations. After the dot-product/GEMM computations an MPI\_Allreduce with summation combines each processor's contributions. Since Allreduce redundantly stores results on all processors, computing the solution to the subproblem and updating all vectors can be performed without communication. Finally, this process is repeated until a termination criterion is met. The mathematical details (and complexity) of solving the subproblem and updating vectors might vary based on the optimization problem (proximal least-squares, SVM, etc.). However, the parallel BCD methods we consider in this paper can be summarized by Figure \ref{fig:cdag}. Each iteration requires synchronization, therefore, avoiding these synchronization costs could lead to faster and more scalable BCD methods for proximal least-squares and SVM.
\paragraph{Main Contributions}
We present synchronization-avoiding (SA) derivations of existing accelerated and non-accelerated CD/BCD through mathematical re-formulation. 
We show that our methods avoid synchronization for $s$ iterations {\it without} altering the numerical stability or convergence behavior in exact arithmetic. We implement the SA-methods in C++ using MPI \cite{mpi} and show that our methods perform $1.2\times$ - $5.1\times$ faster and are more scalable on up to $12$k cores of a Cray XC30 supercomputer on LIBSVM \cite{cc01} datasets.

The rest of the paper is organized as follows: Section \ref{sec:rwork} discusses related approaches and how ours differs. Section \ref{sec:deriv} derives SA-Lasso and shows experimental results in Section IV. Section \ref{sec:deriv_svm} derives SA-SVM and shows experimental results in Section \ref{sec:exp_svm}. Finally, we summarize and discuss future work in Section \ref{sec:conc}.\label{sec:intro}
\section{Related Work}
\paragraph{Communication-Avoiding in Linear Algebra}
This paper extends the $s$-step iterative methods work in NLA and subsequent work on CA iterative linear algebra surveyed by Ballard et. al. \cite{ballard14}. We extend these results to widely used and important problems in ML, namely proximal least-squares and SVM. Our work expands upon and generalizes CA-NLA results to the field of convex optimization.

\paragraph{Communication-Efficient Machine Learning}
Recent work along these lines, like proxCoCoA+ \cite{proxcocoa} propose frameworks that reduce the communication bottleneck. ProxCoCoA+ perform computation on only locally stored data and then combine contributions from all processors. The communication benefits of proxCoCoA+ are inherited from the iteration complexity of performing Newton-type steps on locally stored data.

DUAL-LOCO \cite{dualloco} introduces a framework which reduces communication between processors by communicating a small matrix of randomly projected features. While DUAL-LOCO requires just one communication round, it does not apply to proximal least-squares problems (i.e., Lasso, elastic-net, etc.) and introduces an (additive) approximation error.

CA-SVM \cite{you15} eliminates communication in SVM by performing an initial $K$-means clustering as a pre-processing step to partition the data and subsequently training SVM classifiers locally on each processor. Communication is reduced significantly, but at the cost of accuracy. Like proxCoCoA+, CA-SVM uses a local SVM solver which can be replaced with our SA-variant.

P-packSVM \cite{ppacksvm} applies a similar approach to ours and derives a SA version of SVM using Stochastic Gradient Descent (SGD) as the optimization method. Our work extends this SA approach further to different methods (accelerated and non-accelerated CD/BCD) and extend it to other non-linear optimization problems (proximal least-squares).

Devarakonda et al. \cite{dev16} derive SA-variants of primal and dual BCD methods for L2 regularized least-squares. Our work extends their results to non-linear problems and to sparse convex optimization.

Soori et al. \cite{soori17} derive CA-variants of novel stochastic FISTA (SFISTA) and stochastic Newton (SPNM) methods for the proximal least squares problem. They illustrate that standard loop unrolling techniques can be applied to SFISTA and SPNM to obtain communication-avoiding variants. One the other hand, our work solves the proximal least-squares and SVM problems on accelerated and non-accelerated BCD methods. Furthermore, the synchronization-avoiding technique is adapted from CA-Krylov \cite{ballard14} and $s$-step Krylov subspace methods \cite{chronopoulos89a,chronopoulos89b,chronopoulos96,kim92} work. 

Our SA technique derives alternate forms of existing proximal least-squares and SVM methods by re-arranging the computations to obtain $s$ solution updates per communication round. This allows us to obtain an algorithm whose convergence behavior and sequence of solution updates are equivalent to the original algorithm.
\begin{algorithm}[t!]
\caption{Accelerated Block Coordinate Descent (accBCD) for Lasso}\label{alg:bcd}
\begin{algorithmic}[1]
\State \textbf{Input}: $A \in \mathbb{R}^{m \times n}, b \in \mathbb{R}^m$, $H>1$, $y_0 \in\mathbb{R}^{n}$, $z_0 \in\mathbb{R}^{n}$, $\lambda \in \mathbb{R}$, $\mu  \in \mathbb{Z}_+$ s.t. $\mu \leq n$\\
$\theta_0 = \mu/n,~\tilde y_0 = Ay_0,~\tilde z_0 = Az_0 - b$\\
$q = \ceil{n/\mu}$
\For {$h=1,2,\cdots,H$}
\State choose $\{i_l \in [n] | l = 1, 2, \ldots, \mu \}$ uniformly at random without replacement.
\State $\mathbb{I}_h = \left[e_{i_1}, e_{i_2}, \cdots, e_{i_\mu}\right]$\vspace{0.1cm}
\State Let $\mathbb{A}_h = A\mathbb{I}_h$
\Statex {\bf Communication: Lines 8 and 9.}
\State $G = \mathbb{A}_h^T\mathbb{A}_h$
\State $r_h = \mathbb{A}_{h}^T\left(\theta_{h-1}^2\tilde y_{h-1} + \tilde z_{h-1}\right)$\label{ln:res}
\State $v = $ largest eigenvalue of $G$
\State $\eta_h = \frac{1}{q\theta_{h-1}e_h^Tv}$
\State $g_h = \mathbb{I}_{i_h}^Tz_{h-1} - \eta_hr_h$\label{ln:gradupdate}
\State $\Delta z_h = S_{\lambda\eta_h}(g_h) - \mathbb{I}_{h}^Tz_{h-1}$\label{ln:soln}
\State $z_{h} = z_{h-1} + \mathbb{I}_{h}\Delta z_h$\label{ln:zupdate}
\State $\tilde z_h = \tilde z_{h-1} + \mathbb{A}_{h}\Delta z_h$\label{ln:tzupdate}
\State $y_{h} = y_{h-1} - \frac{1 - q\theta_{h-1}}{\theta_{h-1}^2}\mathbb{I}_{h}\Delta z_h$\label{ln:yupdate}
\State $\tilde y_h = \tilde y_{h-1} - \frac{1 - q\theta_{h-1}}{\theta_{h-1}^2}\mathbb{A}_{h}\Delta z_h$\label{ln:tyupdate}
\State $\theta_h = \frac{\sqrt{\theta_{h-1}^4 + 4\theta_{h-1}^2} - \theta_{h-1}^2}{2}$
\EndFor
\State \textbf{Output} $\theta_H^2y_H + z_H$
\end{algorithmic}
\end{algorithm}\label{sec:rwork}
\section{Synchronization-Avoiding LASSO}
In this section, we derive a SA version of the accelerated block coordinate descent (accBCD) algorithm for the Lasso problem. The derivation of SA-accBCD (Synchronization-Avoiding accBCD) relies on unrolling the vector update recurrences $s$ times and re-arranging the updates and dependent computations to avoid synchronization.

Given a matrix $A \in \mathbb{R}^{m \times n}$ with $m$ data points and $n$ features, a vector of labels $b \in \mathbb{R}^{m}$, and regularization parameter $\lambda \in \mathbb{R}$, the Lasso problem aims to find the solution $x \in \mathbb{R}^n$ that solves the optimization problem:
\begin{align*}
\argmin_{x \in \mathbb{R}^n} \frac{1}{2}\|Ax - b\|^2_2 + \lambda ||x||_1
\end{align*}
This problem can be solved with many iterative algorithms. We consider the accelerated BCD (accBCD) algorithm described in \cite[Algorithm 2]{fercoq15} (see Algorithm \ref{alg:bcd} in this paper). Nesterov's acceleration \cite{nesterov05} is adapted to BCD\cite{fercoq15} through the introduction of additional vectors ($y_h, z_h, \tilde y_h,$ and $\tilde z_h$) and scalar ($\theta_h$) updates. The solution vector at each iteration $h$ is implicitly computed since $x_h = \theta_h y_h + z_h$, but need not be explicitly computed until termination. Aside from complications due to acceleration, the remaining computations follow the BCD steps illustrated in Figure \ref{fig:cdag}.

We assume that $A$ is 1D-row partitioned so partial dot-products (lines 8 and 9 in Alg. \ref{alg:bcd}) can be computed on each processor. Lines $4$-$7$ in Alg. \ref{alg:bcd} randomly select $\mu$ column indices and extract them from $A$. Lines $8$-$9$ in Alg. \ref{alg:bcd} compute dot-products. Lines $11$-$17$ Alg. \ref{alg:bcd} compute the solution to the subproblem and update scalars and vectors for the next iteration. We assume that $H$ iterations of the algorithm are performed (where $H$ depends on the termination criterion).
Line 12 in Alg. \ref{alg:bcd} applies the soft-thresholding function defined in \eqref{eq:sth} required for Lasso.
At each iteration we compute $v$, the optimal Lipschitz constant, which is the largest eigenvalue of the, small, $\mu \times \mu$ Gram matrix (line 10 in Alg \ref{alg:bcd}). Since $G$ is replicated on all processors (after the MPI\_Allreduce \cite{mpi}), line 10 does not require communication. An approximate Lipschitz constant can be used but we compute the optimal constant for fast convergence.
The recurrences in lines \ref{ln:res}--\ref{ln:tyupdate} of Alg. \ref{alg:bcd} can be unrolled to avoid synchronization. We begin the SA derivation by changing the loop index from $h$ to $sk + j$ where $k$ is the outer loop index, $s$ is the recurrence unrolling parameter, and $j$ is the inner loop index. Let us assume that we are at iteration $sk + 1$ and have just computed the vectors $z_{sk}$, $\tilde z_{sk}$, $y_{sk}$, and $\tilde y_{sk}$. From this $\Delta z_{sk + 1}$ can be computed~~by\footnote{We ignore scalar updates since they can be redundantly stored and computed on all processors.}

\begin{align*}
r_{sk + 1} &= \mathbb{A}_{{sk + 1}}^T\left(\theta_{sk}^2\tilde y_{sk} + \tilde z_{sk}\right),\\
g_{sk + 1} &= \mathbb{I}_{{sk  + 1}}^Tz_{sk} - \eta_{sk + 1}r_{sk + 1},\\
\Delta z_{sk + 1} &= S_{\lambda\eta_{sk + 1}}(g_{sk + 1}) - \mathbb{I}_{{sk + 1}}^Tz_{sk}.
\end{align*}
\begin{algorithm}[t!]
\caption{Synchronization-Avoiding Accelerated Block Coordinate Descent (SA-accBCD) for Lasso}\label{alg:cabcd}
\begin{algorithmic}[1]
\State \textbf{Input}: $A \in \mathbb{R}^{m \times n}, b \in \mathbb{R}^m$, $H>1$, $y_0 \in\mathbb{R}^{n}$, $z_0 \in\mathbb{R}^{n}$, $\lambda \in \mathbb{R}$, $\mu  \in \mathbb{Z}_+$ s.t. $\mu \leq n$
\State $\theta_0 = \mu/n,~\tilde y_0 = Ay_0,~\tilde z_0 = Az_0 - b$
\State $q = \ceil{n/\mu}$
\For {$k=1,2,\cdots,\frac{H}{s}$}
  \For{$j = 1, 2, \cdots, s$}
  \State choose $\{i_l \in [n] | l = 1, 2, \ldots, \mu \}$ uniformly at 
  \Statex \quad \quad \quad random without replacement.
  \State $\mathbb{I}_{sk+j} = \left[e_{i_1}, e_{i_2}, \cdots, e_{i_\mu}\right]$\vspace{0.1cm}
  \State Let $\mathbb{A}_{sk + j} = A\mathbb{I}_{sk+j} $
  \State $\theta_{sk + j} = \frac{\sqrt{\theta_{sk + j-1}^4 + 4\theta_{sk + j-1}^2} - \theta_{sk + j-1}^2}{2}$
  \EndFor
  \Statex {\bf Communication: Lines 11 and 12.}
  \State Let $Y = \begin{bmatrix}\mathbb{A}_{sk + 1}, \mathbb{A}_{sk + 2}, \cdots, \mathbb{A}_{sk + s}\end{bmatrix}$.
  \State $G = Y^TY$.
  \State $\begin{bmatrix}\tilde y'_{sk + 1} & \tilde y'_{sk+2} & \ldots & \tilde y'_{sk + s}\\ \tilde z'_{sk + 1} & \tilde z'_{sk+2} & \ldots & \tilde z'_{sk + s}\end{bmatrix}^T = Y^T \left[\tilde y_{sk}\quad \tilde z_{sk} \right]$.
  \For{$j = 1, 2, \cdots, s$}
  \State $v = $ large eigenvalue of $\mathbb{A}_{sk + j}^T\mathbb{A}_{sk + j}$.
  \State $\eta_{sk + j} = \frac{1}{q\theta_{sk + j -1}e_{sk + j}^Tv}$
  \State Compute $r_{sk + j}$ by equation\footnotemark \eqref{eq:res} 
  \State Compute $g_{sk + j}$ by equation \eqref{eq:grad} 
 \State Compute $\Delta z_{sk + j}$ by equation \eqref{eq:delz}
  \State $z_{sk + j} = z_{sk + j-1} + \mathbb{I}_{{sk + j}}\Delta z_{sk + j}$
  \State $\tilde z_{sk + j} = \tilde z_{sk + j-1} + \mathbb{A}_{{sk + j}}\Delta z_{sk + j}$
  \State $y_{sk + j} = y_{sk + j-1} - \frac{1 - q\theta_{sk + j -1}}{\theta_{sk + j-1}^2}\mathbb{I}_{{sk + j}}\Delta z_{sk + j}$
  \State $\tilde y_{sk + j} = \tilde y_{sk + j -1} - \frac{1 - q\theta_{sk + j - 1}}{\theta_{sk + j -1}^2}\mathbb{A}_{{sk + j}}\Delta z_{sk + j}$
  \EndFor
\EndFor
\State \textbf{Output} $\theta_H^2y_H + z_H$
\end{algorithmic}
\end{algorithm}
By unrolling the vector update recurrences for $z_{sk + 1}$, $\tilde y_{sk + 1}$, and $\tilde z_{sk + 1}$ (lines \ref{ln:zupdate}, \ref{ln:tzupdate}, and \ref{ln:tyupdate}), we can compute $r_{sk + 2}$, $g_{sk + 2}$, and $\Delta z_{sk + 2}$ in terms of $z_{sk}$, $\tilde z_{sk}$, and $\tilde y_{sk}$
\begin{align*}
\begin{split}
  r_{sk + 2} &= \mathbb{A}_{{sk + 2}}^T\biggr(\theta_{sk + 1}^2\tilde y_{sk} - \theta_{sk + 1}^2\frac{1 - q\theta_{sk}}{\theta_{sk}^2}\mathbb{A}_{{sk + 1}}\Delta z_{sk + 1} \\&+ \tilde z_{sk} + \mathbb{A}_{{sk + 1}}\Delta z_{sk + 1}\biggr),
  \end{split}\\
\begin{split}
  &= \theta_{sk + 1}^2\mathbb{A}_{{sk + 2}}^T\tilde y_{sk} + \mathbb{A}_{{sk + 2}}^T\tilde z_{sk} \\&- \left(\theta_{sk + 1}^2 \frac{1 - q\theta_{sk}}{\theta_{sk}^2}- 1\right)\mathbb{A}_{{sk + 2}}^T\mathbb{A}_{{sk + 1}}\Delta z_{sk + 1},
\end{split}\\
  g_{sk + 2} &= \mathbb{I}_{{sk  + 2}}^Tz_{sk} + \mathbb{I}_{{sk  + 2}}^T\mathbb{I}_{{sk  + 1}}\Delta z_{sk + 1} - \eta_{sk + 2}r_{sk + 2},\\
  \Delta z_{sk + 2} &= S_{\lambda\eta_{sk + 2}}(g_{sk + 2}) - \mathbb{I}_{{sk + 2}}^Tz_{sk} - \mathbb{I}_{{sk  + 2}}^T\mathbb{I}_{{sk  + 1}}\Delta z_{sk + 1}.
\end{align*}
By induction we can show that $r_{sk + j}$, $g_{sk + j}$, and $\Delta z_{sk + j}$ can be computed in terms of $z_{sk}$, $\tilde z_{sk}$, and $\tilde y_{sk}$
\begin{align}
\begin{split}
r_{sk + j} &= \theta_{sk + j - 1}^2\mathbb{A}_{{sk + j}}^T\tilde y_{sk}+ \mathbb{A}_{{sk + j}}^T\tilde z_{sk}\\
  &\hspace{-1cm}- \sum_{t = 1}^{j-1}\left(\theta_{sk + j - 1}^2\frac{1 - q\theta_{sk + t - 1}}{\theta_{sk + t - 1}^2} - 1\right)\mathbb{A}_{{sk + j}}^T\mathbb{A}_{{sk + t}}\Delta z_{sk + t}\label{eq:res}
\end{split}\\
  g_{sk + j} &= \mathbb{I}_{{sk  + j}}^Tz_{sk} - \eta_{sk + j}r_{sk + j}  + \sum_{t = 1}^{j-1}\mathbb{I}_{{sk  + j}}^T\mathbb{I}_{{sk  + t}}\Delta z_{sk + t}\label{eq:grad}\\
  \begin{split}
  \Delta z_{sk + j} &= S_{\lambda\eta_{sk + j}}(g_{sk + j})- \mathbb{I}_{{sk + j}}^Tz_{sk} \\
  &- \sum_{t = 1}^{j-1}\mathbb{I}_{{sk  + j}}^T\mathbb{I}_{{sk  + t}}\Delta z_{sk + t}\label{eq:delz}
  \end{split}
\end{align}
\footnotetext{Since $\mathbb{A}^T_{sk +j }\tilde y_{sk} = \tilde y'_{sk}$ and $\mathbb{A}^T_{sk +j }\tilde y_{sk} = \tilde z'_{sk}$, no additional computation or communication is needed to form those vectors. }
\begin{table*}[t!]
\begin{center}
\begin{tabular}{c| c | c | c | c}
\hline
\multicolumn{5}{c}{Summary of theoretical costs}\\ \hline \hline
Algorithm & Ops cost (F) & Memory cost (M) & Latency cost (L) & Message Size cost (W)\\ \hline
{accBCD} & {$O\left(\frac{H\mu^2fm}{P} + H\mu^3\right)$} & {$O\left(\frac{fmn + m}{P} + \mu^2 + n\right)$} & $O\left(H\log P\right)$& $O\left(H\mu^2\log P\right)$\\ \cline{1-1} \hline
	{SA-accBCD} & {$O\left(\frac{H\mu^2\bm{s}fm}{P} + H\mu^3\right)$} & $O\left(\frac{fmn + m}{P} + \mu^2\bm{s^2} + n\right)$ & $O\left(\frac{H}{\bm{s}}\log P\right)$& $O\left(H\bm{s}\mu^2\log P\right)$\\ \hline
\end{tabular}
\end{center}
\caption{Ops (F), Latency (L), Bandwidth (W) and Memory per processor (M) costs comparison along the critical path of classical accBCD and SA-accBCD. $H$ is the number of iterations and we assume that $A \in \mathbb{R}^{m \times n}$ is sparse with $fmn$ non-zeros that are uniformly distributed, $0 < f \leq 1$ is the density of $A$ (i.e. $f = \frac{nnz(A)}{mn}$), $P$ is the number of processors and $s$ is the recurrence unrolling parameter. $f\mu m$ is the non-zeros of the $\mu \times m$ matrix with $\mu$ sampled columns from $A$ at each iteration. We assume that the $\mu \times \mu$ and Gram matrix computed at each iteration are dense.}
\label{tbl:sumres}
\end{table*}
for $j = 1, 2, \ldots, s$. Notice that due to the recurrence unrolling we can defer the updates to $z_{sk}$, $y_{sk}$, $\tilde z_{sk}$, and $\tilde y_{sk}$ for $s$ iterations. The summation in \eqref{eq:res} computes the Gram-like matrices, $\mathbb{A}_{{sk + j}}^T\mathbb{A}_{{sk + t}}$, for $t = 1, 2, \cdots, j - 1$. Synchronization can be avoided in these computations by computing the $s\mu \times s\mu$ Gram matrix $G = [\mathbb{A}_{{sk + 1}}, \mathbb{A}_{{sk + 2}}, \cdots, \mathbb{A}_{{sk + s}}]^T[\mathbb{A}_{{sk + 1}}, \mathbb{A}_{{sk + 2}}, \cdots, \mathbb{A}_{{sk + s}}]$ once before the inner loop\footnote{$G$ is symmetric so computing just the upper/lower triangular part reduces flops and message size by $2\times$.} and redundantly storing it on all processors. Synchronization can be avoided in the summation in \eqref{eq:grad} by initializing the random number generator on all processors to the same seed. Finally, at the end of the $s$ inner loop iterations we can perform the vector updates
\begin{align}
  z_{sk + s} &= z_{sk} + \sum_{t = 1}^{s}\mathbb{I}_{i_{sk + t}}\Delta z_{sk + t}\\
  \tilde z_{sk + s} &= \tilde z_{sk} + \sum_{t = 1}^{s}\mathbb{A}_{{sk + t}}\Delta z_{sk + t}\\
  y_{sk + s} &= y_{sk} - \sum_{t = 1}^{s}\frac{1 - q\theta_{sk + t-1}}{\theta_{sk + t-1}^2}\mathbb{I}_{{sk + t}}\Delta z_{sk + t}\\
  \tilde y_{sk + s} &= \tilde y_{sk} - \sum_{t = 1}^{s}\frac{1 - q\theta_{sk + t-1}}{\theta_{sk + t-1}^2}\mathbb{A}_{{sk + t}}\Delta z_{sk + t}   .
\end{align}

The resulting Synchronization-Avoiding accBCD (SA-accBCD) algorithm is shown in Algorithm \ref{alg:cabcd}. Since our SA technique relies on rearranging the computations, the convergence rates and behavior of the standard accelerated BCD algorithm (Alg. \ref{alg:bcd}) is the same (in exact arithmetic).

SA-accBCD computes a larger $s\mu \times s\mu$ Gram matrix every $s$ iterations, which results in a computation-communication tradeoff where SA-accBCD increases the flops and message size in order to reduce the latency by $s$. If the latency cost is the dominant term then SA-accBCD can attain $s$-fold speedup over accBCD. In general there exists a tradeoff between $s$ and the speedups attainable. Table \ref{tbl:sumres} summarizes the operational, storage, and communication costs of the SA and non-SA methods.
\label{sec:deriv}
\section{Experimental Results: LASSO}
\begin{table}
\begin{center}
\footnotesize
\begin{tabular}{l|c|c|c}
\hline
\multicolumn{4}{c}{Summary of datasets}\\
\hline\hline
\multicolumn{1}{c}{Name} &  \multicolumn{1}{|c|}{Features} &Data Points & \multicolumn{1}{|c}{NNZ$\%$}\\
\hline
url & $3,231,961$ & $2,396,130$ & $0.0036$\\\hline
news20 & $62,061$ &$15,935$ & $0.13$\\ \hline
covtype & $54$ & $581,012$ & $22$\\ \hline
epsilon & $2,000$ & $400,000$ & $100$\\ \hline
leu &$7,129$ & $38$ & $100$\\ \hline
\end{tabular}
\end{center}
\caption{Properties of the LIBSVM datasets used in our numerical and performance experiments.}
\label{tbl:dsets}
\end{table}
In this section we present experimental results for our SA variant and explore the numerical and performance tradeoffs. The recurrence unrolling we propose requires computation of Gram-like matrices whose condition numbers may adversely affect the numerical stability of SA-accBCD. We also re-order the sequence of updates of the solution and residual vectors, which could also lead to numerical instability. We begin in Section \ref{sec:numexp} with experiments that illustrate that SA-accBCD is numerically stable.
\subsection{Convergence behavior}\label{sec:numexp}
\begin{figure}[t!]
  \centering
\begin{subfigure}{.235\textwidth}
\centering
\includegraphics[trim = 0.1in 2.5in 0.1in 2.5in, clip,width=1\textwidth]{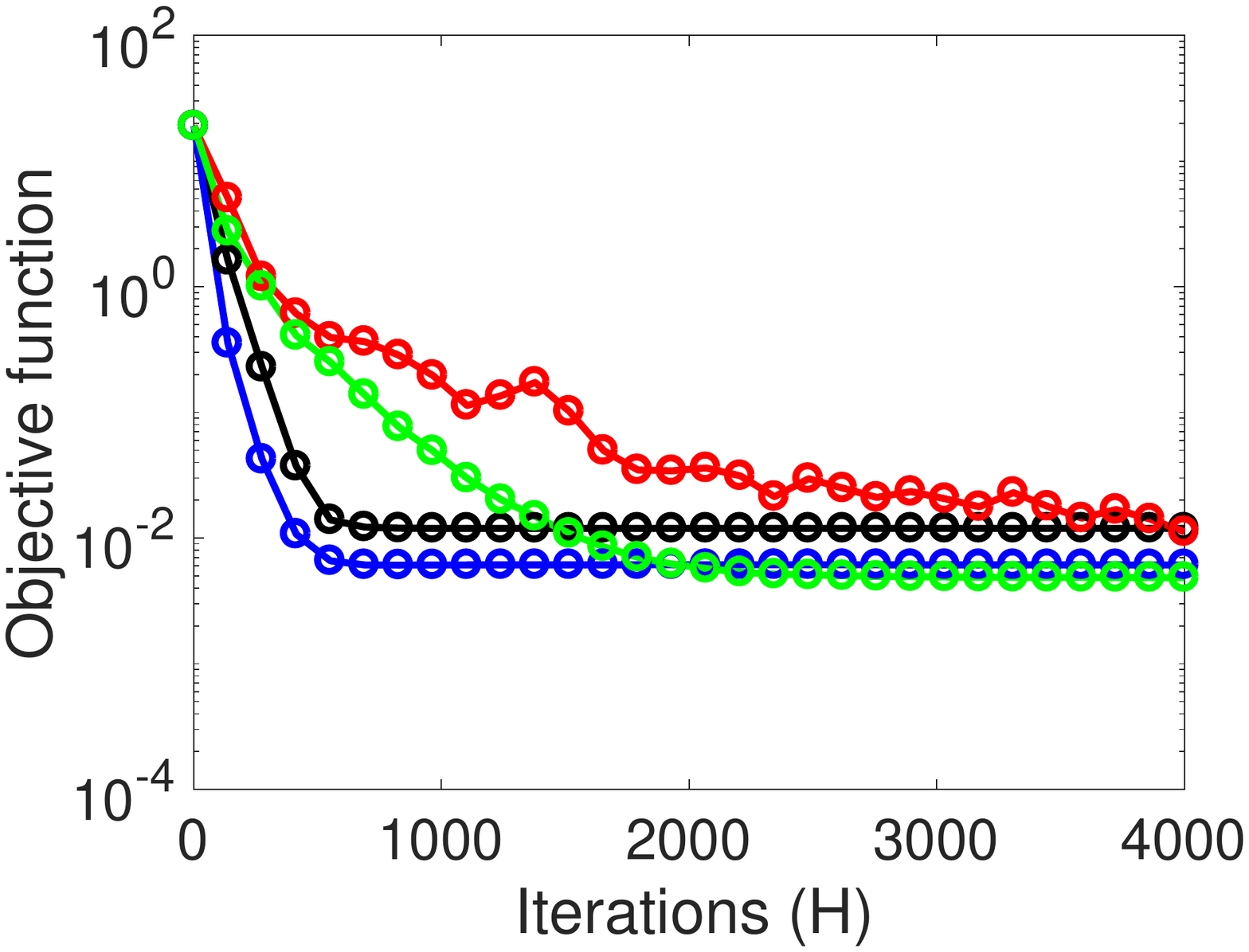}
\caption{leu}
\label{fig:leuobj}
\end{subfigure}
\begin{subfigure}{.235\textwidth}
\centering
\includegraphics[trim = 0.1in 2.5in 0.1in 2.5in, clip,width=1\textwidth]{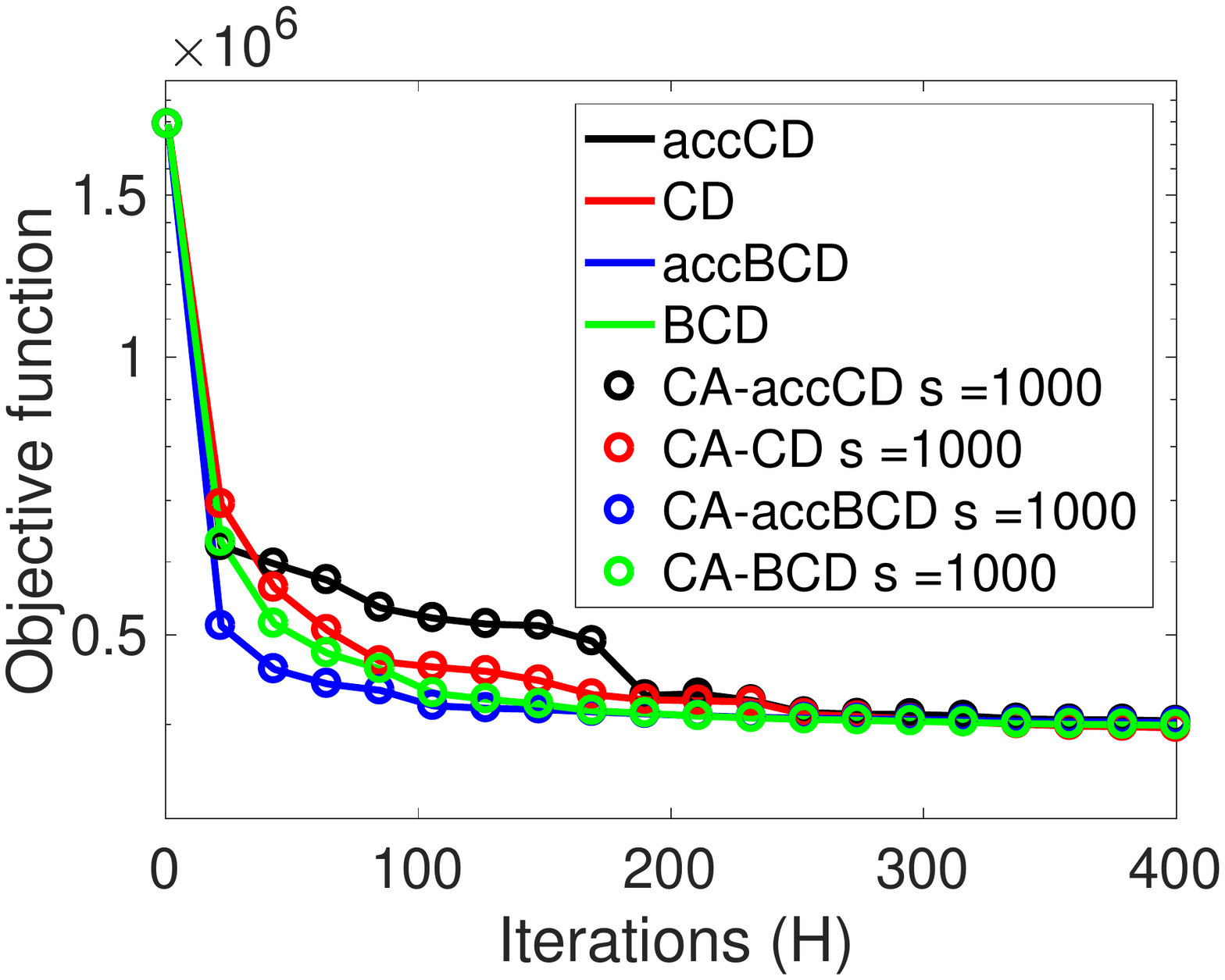}
\caption{covtype}
\label{fig:covtypeobj}
\end{subfigure}

\begin{subfigure}{.235\textwidth}
\centering
\includegraphics[trim = 0.1in 2.5in 0.1in 2.5in, clip,width=1\textwidth]{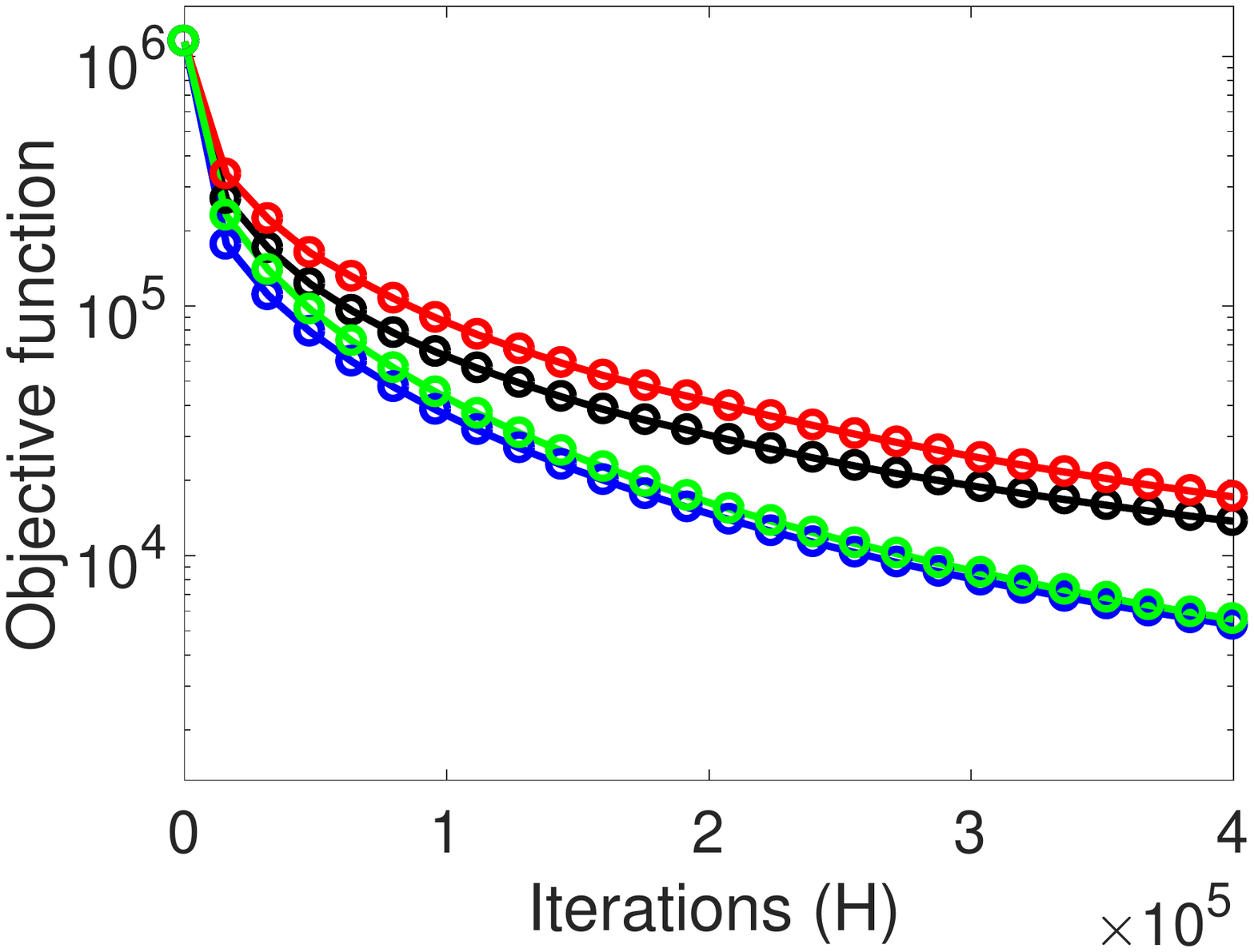}
\caption{news20}
\label{fig:news20obj}
\end{subfigure}
\caption{We compare the convergence of accCD, CD, accBCD, BCD against their SA variants (with $s = 1000$). $\lambda = 100\sigma_{min}$, where $\sigma_{min}$ is the smallest singular value. $\mu = 1$ for accCD and CD. $\mu = 8$ for accBCD and BCD.}
\label{fig:conv}
\end{figure}

\begin{figure*}[th!]
\centering
\begin{subfigure}{.24\textwidth}
\centering
\includegraphics[trim = 0.1in 2.5in 0.1in 2.5in, clip,width=1\textwidth]{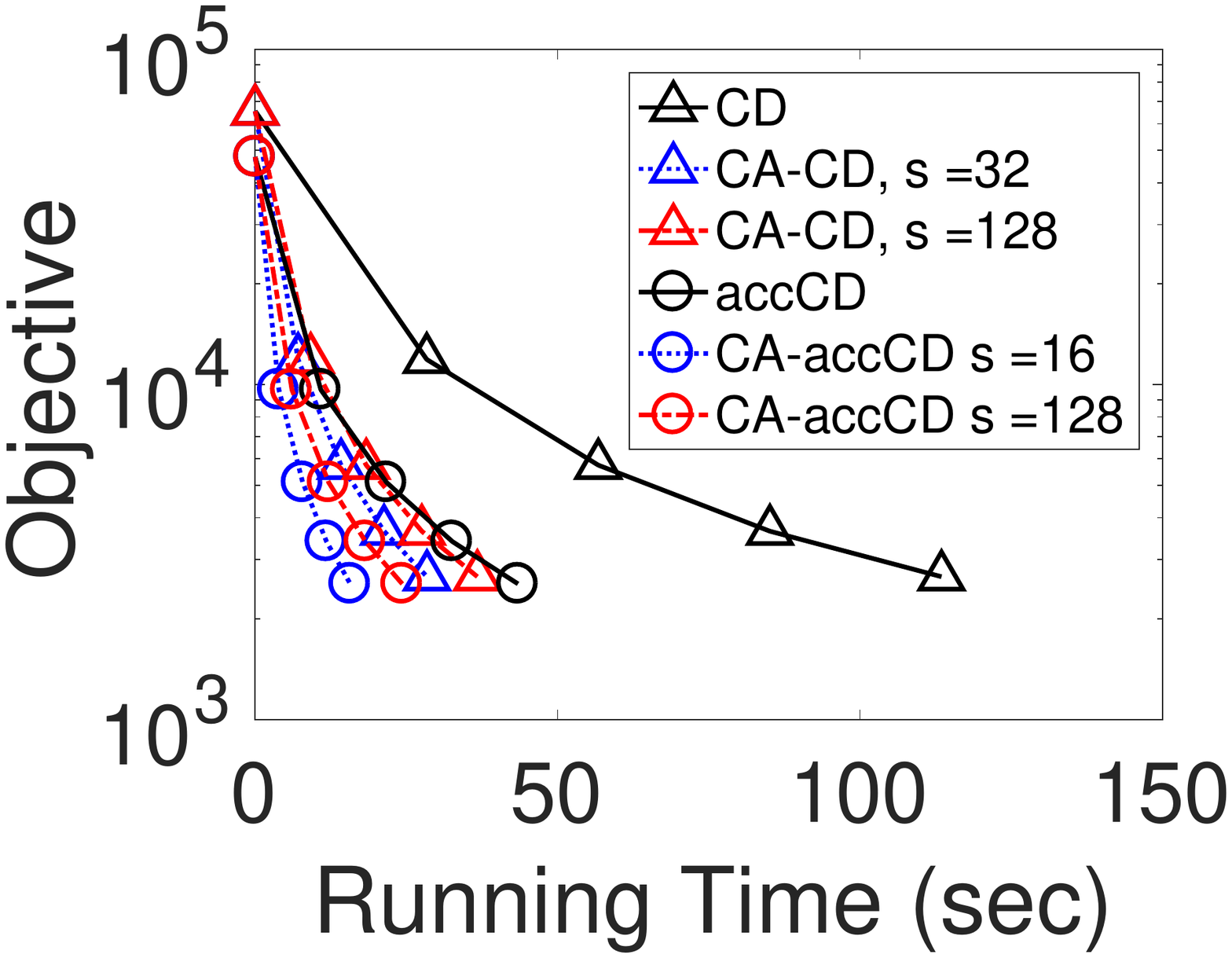}
\label{fig:news20cdobj}
\end{subfigure}
\begin{subfigure}{.24\textwidth}
\centering
\includegraphics[trim = 0.1in 2.5in 0.1in 2.5in, clip,width=1\textwidth]{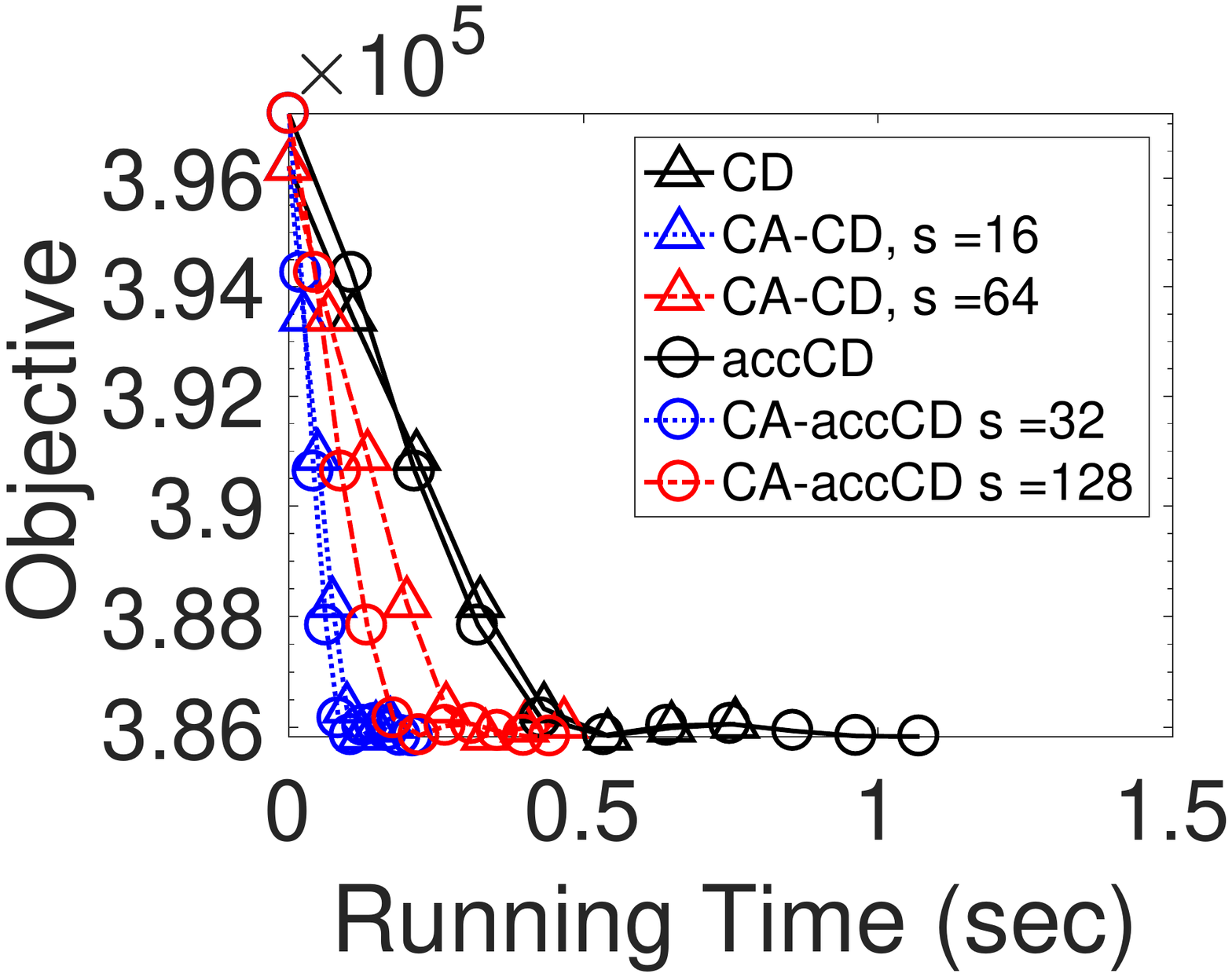}
\label{fig:covtypecdobj}
\end{subfigure}
\begin{subfigure}{.24\textwidth}
\centering
\includegraphics[trim = 0.1in 2.5in 0.1in 2.5in, clip,width=1\textwidth]{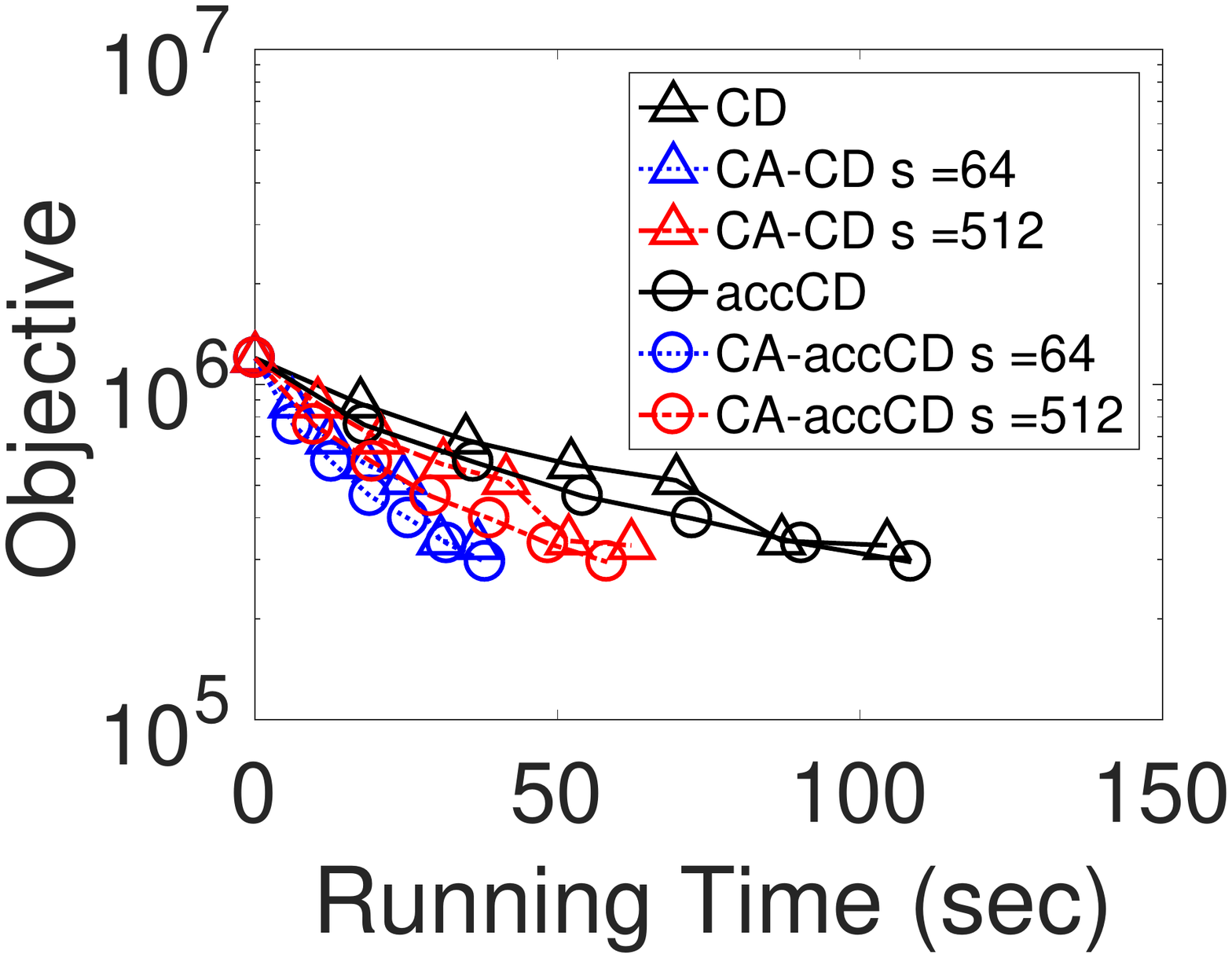}
\label{fig:urlcdobj}
\end{subfigure}
\begin{subfigure}{.24\textwidth}
\centering
\includegraphics[trim = 0.1in 2.5in 0.1in 2.5in, clip,width=1\textwidth]{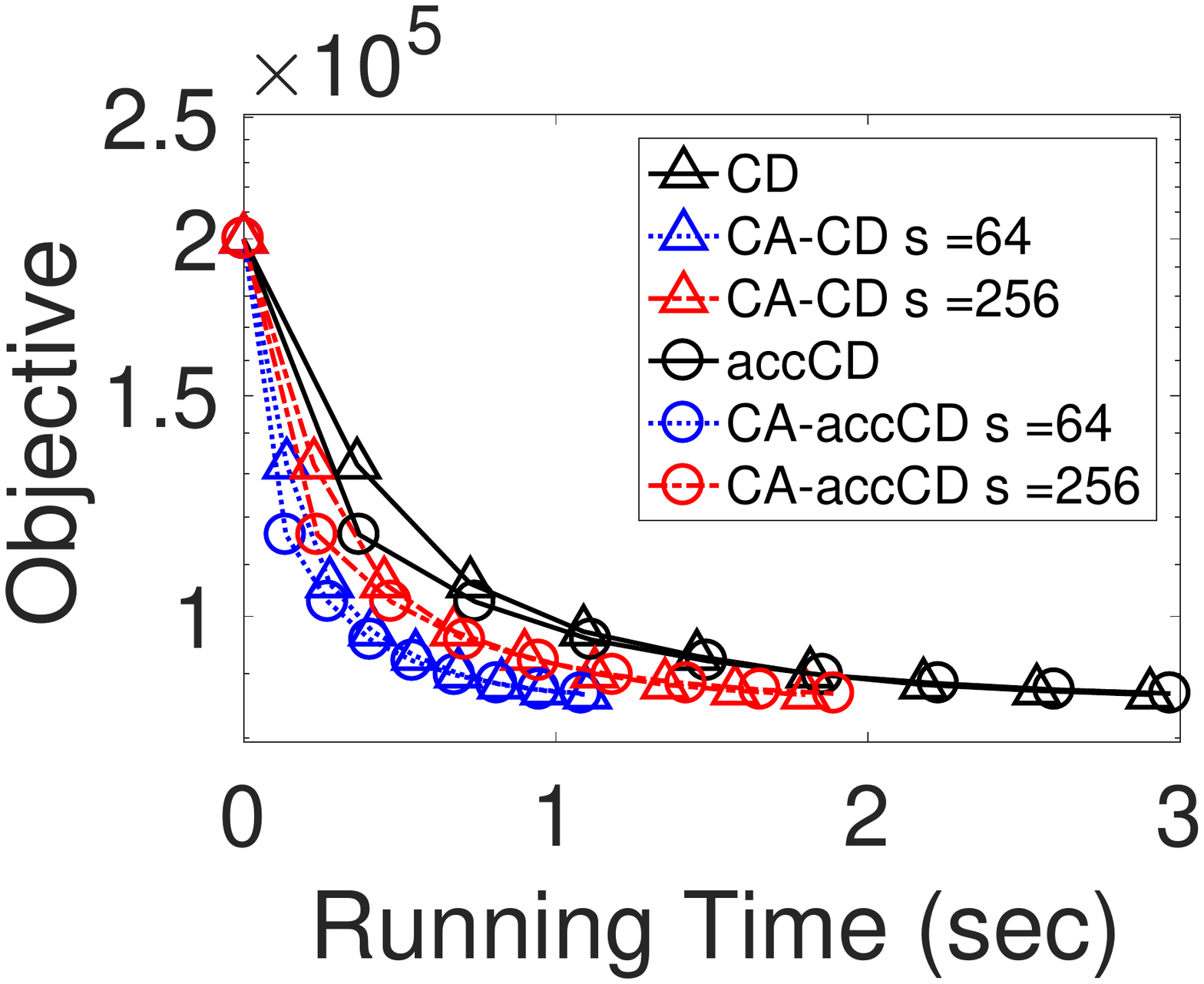}
\label{fig:epsiloncdobj}
\end{subfigure}

\begin{subfigure}{.24\textwidth}
\centering
\includegraphics[trim = 0.1in 2.5in 0.1in 2.5in, clip,width=1\textwidth]{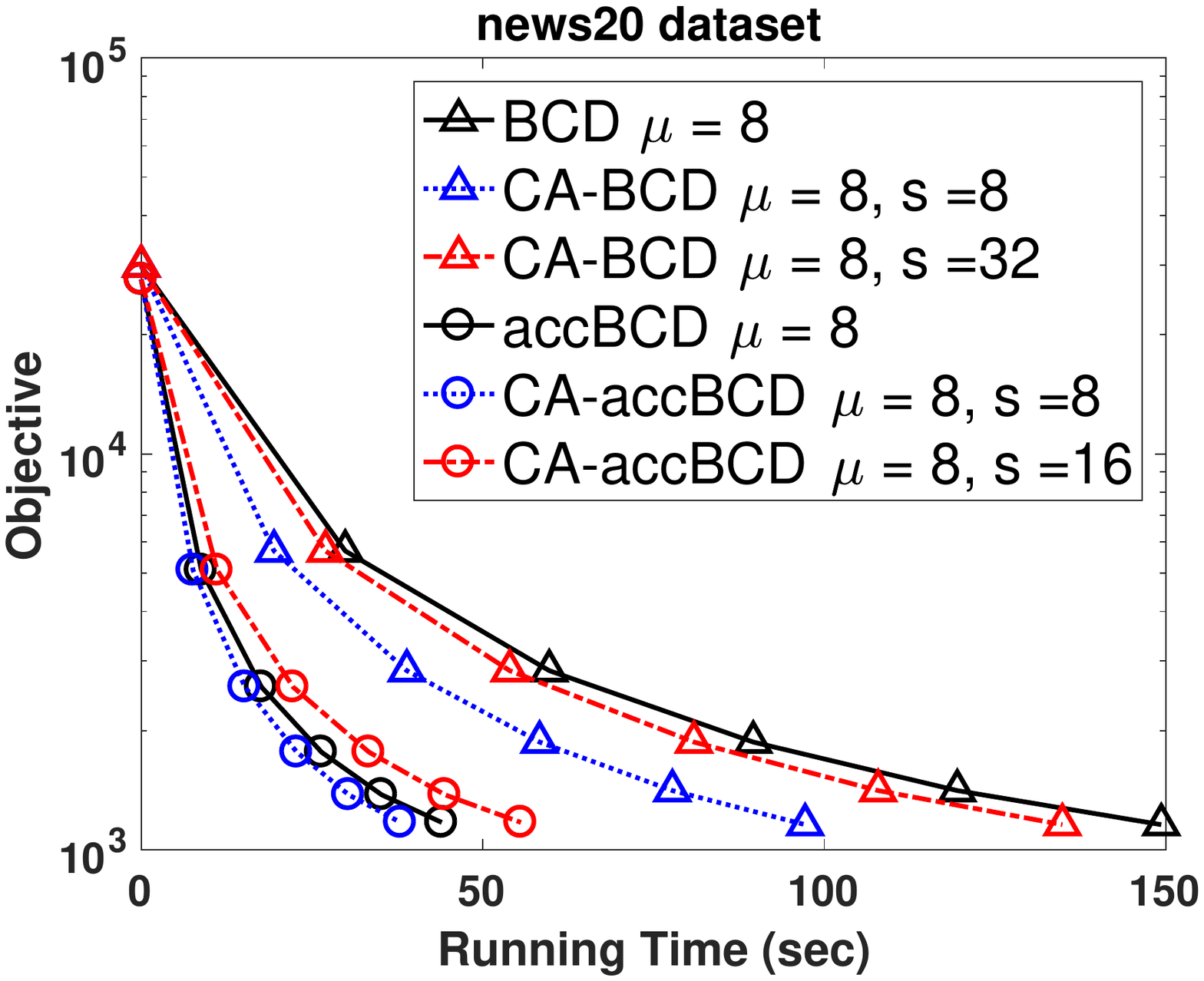}
\label{fig:news20bcdobj}
\caption{news20; Processors = $768$}
\end{subfigure}
\begin{subfigure}{.24\textwidth}
\centering
\includegraphics[trim = 0.1in 2.5in 0.1in 2.5in, clip,width=1\textwidth]{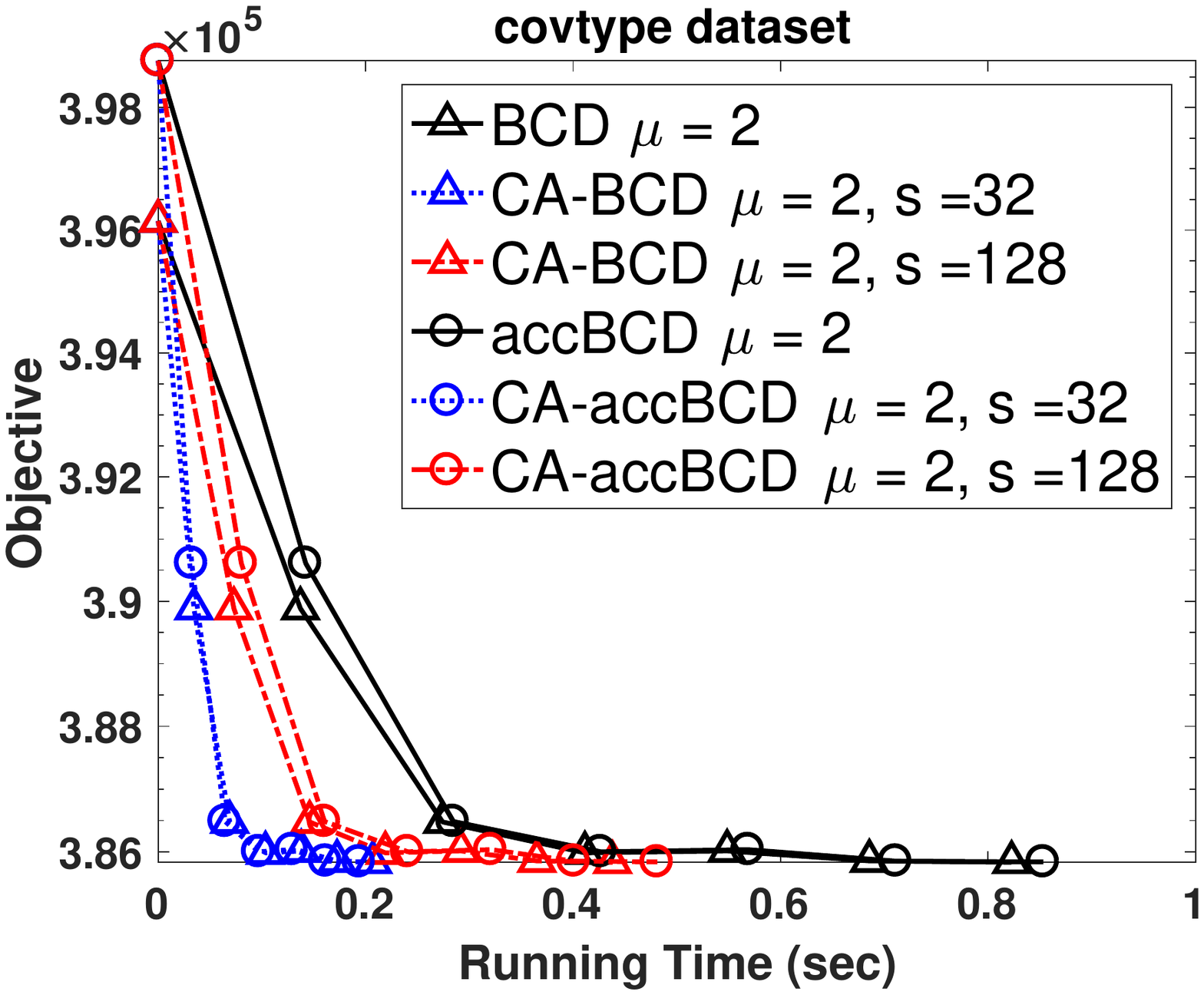}
\label{fig:covtypebcdobj}
\caption{covtype; Processors = $3072$}
\end{subfigure}
\begin{subfigure}{.24\textwidth}
\centering
\includegraphics[trim = 0.1in 2.5in 0.1in 2.5in, clip,width=1\textwidth]{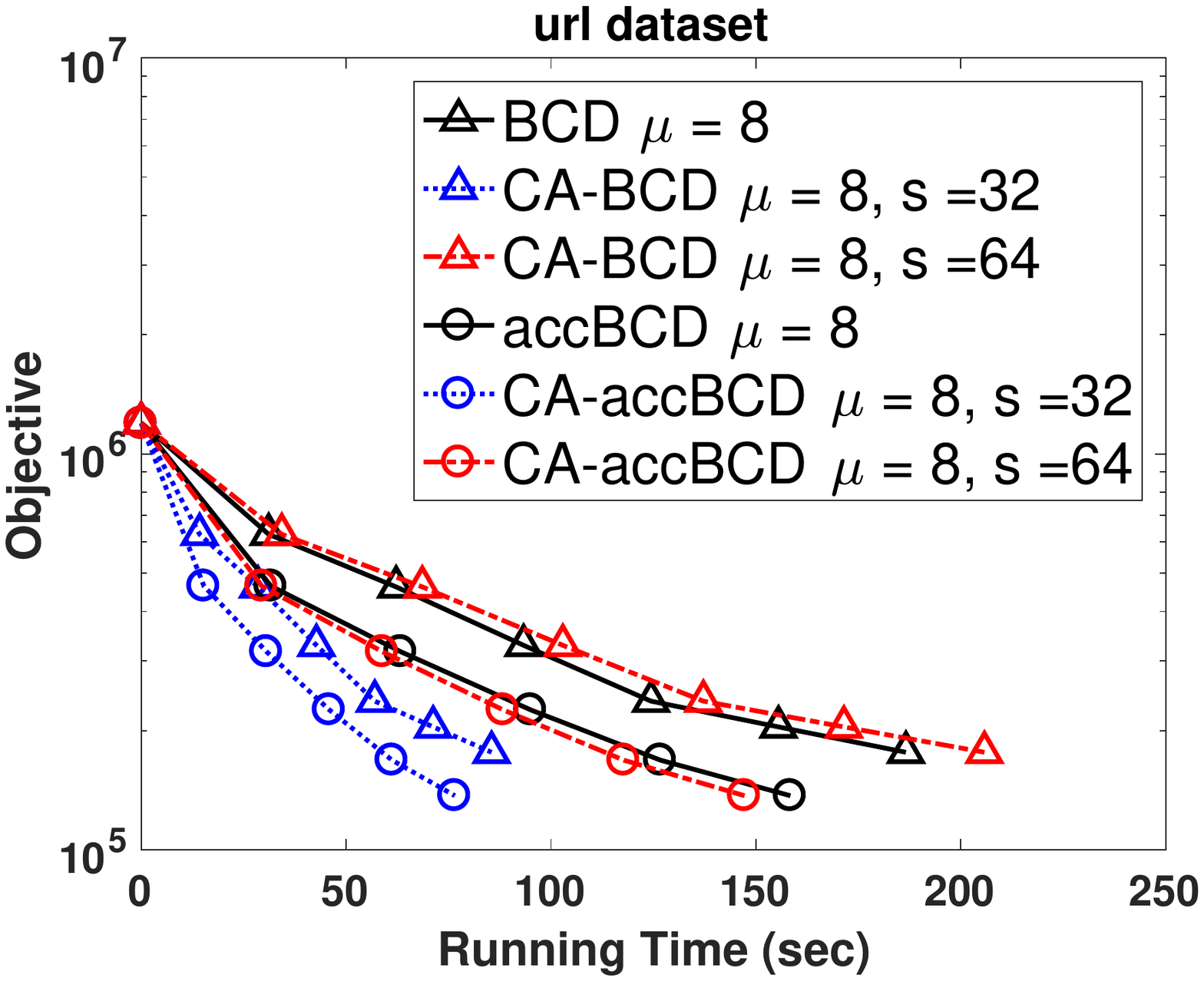}
\label{fig:urlbcdobj}
\caption{url; Processors = $12288$}
\end{subfigure}
\begin{subfigure}{.24\textwidth}
\centering
\includegraphics[trim = 0.1in 2.5in 0.1in 2.5in, clip,width=1\textwidth]{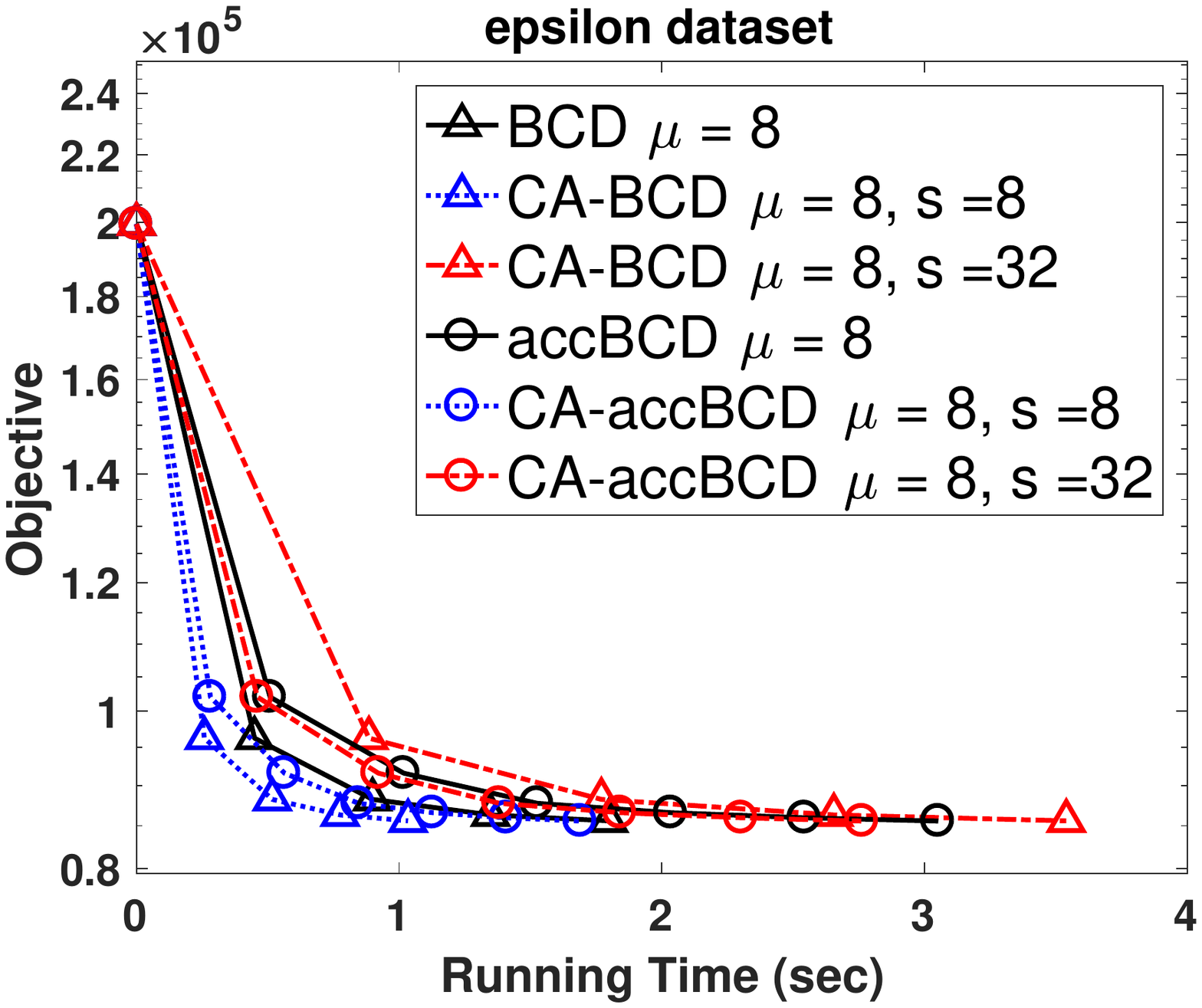}
\label{fig:epsilonbcdobj}
\caption{epsilon; Processors = $12288$}
\end{subfigure}
\caption{We compare SA vs non-SA running times vs. convergence of CD, accCD (top row), BCD, accBCD (bottom row).}
\label{fig:sobj}
\end{figure*}
We explore the tradeoff between convergence behavior, block size, and $s$ (the recurrence unrolling parameter) for the SA-accBCD algorithm and compare it to the behavior of the standard accBCD algorithm. All numerical stability experiments were performed in MATLAB version R2016b on an Intel i7. The datasets used in our experiments were obtained from the LIBSVM repository \cite{cc01} and are summarized in Table \ref{tbl:dsets}. We measure the convergence behavior by plotting the objective function value: $f(A,b,x_h) = \frac{1}{2}\|Ax_h - b\|^2_2 + \lambda \|x_h\|_1$ at each iteration. For all experiments, we set $\lambda = 100\sigma_{min}$. We report the convergence vs. iterations to illustrate any differences in convergence behavior. 

Figure \ref{fig:conv} shows the convergence behavior of the datasets in Table \ref{tbl:dsets} for several blocksizes and with $s = 1000$ for SA-accBCD. The results show that larger blocksizes converge faster than $\mu = 1$, but at the expense of more computation (and larger message sizes in the distributed-memory setting). Comparing SA-accBCD and accBCD we observe no numerical stability issues for $s$ as large as $1000$ for all datasets tested (in theory we can avoid communication for $1000$ iterations). Table \ref{tbl:relobjerr} shows the final relative objective error of the SA-methods compared to the non-SA methods: $\frac{|f(.)_\text{non-SA} - f(.)_\text{SA}|}{f(.)_\text{non-SA}}$. This suggests that the additional computation and message size costs are the performance limiting factors and not numerical instability.
\begin{table}
\begin{center}
\footnotesize
\begin{tabular}{l|c|c|c}
\hline
\multicolumn{4}{c}{Relative objective error}\\
\hline\hline
&\multicolumn{1}{c}{leu} & covtype & \multicolumn{1}{|c}{news20}\\
\cline{2-4}
SA-accCD & 1.3851e-16 & 2.1514e-16 & 6.6324e-17\\\hline
SA-CD & 1.6492e-16& 1.4203e-16& 3.2567e-17\\ \hline
SA-accBCD & 8.2004e-17& 2.2616e-16 &5.6153e-17\\ \hline
SA-BCD & 9.0930e-17& 2.6451e-16 & 8.8625e-17\\ \hline
\end{tabular}
\end{center}
\caption{Final relative objective error of the SA vs. non-SA methods (from Figure \ref{fig:conv}). Machine precision is 2.2e-16.} 
\label{tbl:relobjerr}
\end{table}

\subsection{Performance results}\label{sec:perfexp}

In this section, we present experimental results to show that the SA-methods in Section \ref{sec:numexp} are faster than their non-SA variants. We consider the datasets in Table \ref{tbl:dsets} which were chosen to illustrate performance and speedups on over/under-determined, sparse and dense datasets to illustrate that speedups are independent of those factors.

We implement the algorithms in C++ using the Message Passing Interface (MPI) \cite{mpi} for high-performance, distributed-memory parallel processing. The local linear algebra computations are performed using the Intel MKL library for Sparse and Dense BLAS \cite{lawson79} routines. All methods were tested on a Cray XC30 supercomputer at NERSC which has 24 processors per node and 128GB of memory. The implementation divides the dataset row-wise, however, the SA-methods generalize to other data layout schemes. We choose row-wise since it results in the lowest per iteration communication cost of $O(\log P)$ \cite{dev16}. All datasets are stored using Compressed Sparse Row format (3-array variant). Vectors in $\mathbb{R}^n$ are replicated on all processors and vectors in $\mathbb{R}^m$ are partitioned (similar to $A$).
\begin{figure*}[h]
  \centering
\begin{subfigure}{.24\textwidth}
\centering
\includegraphics[trim = 0.1in 2.5in 0.1in 2.5in, clip,width=1\textwidth]{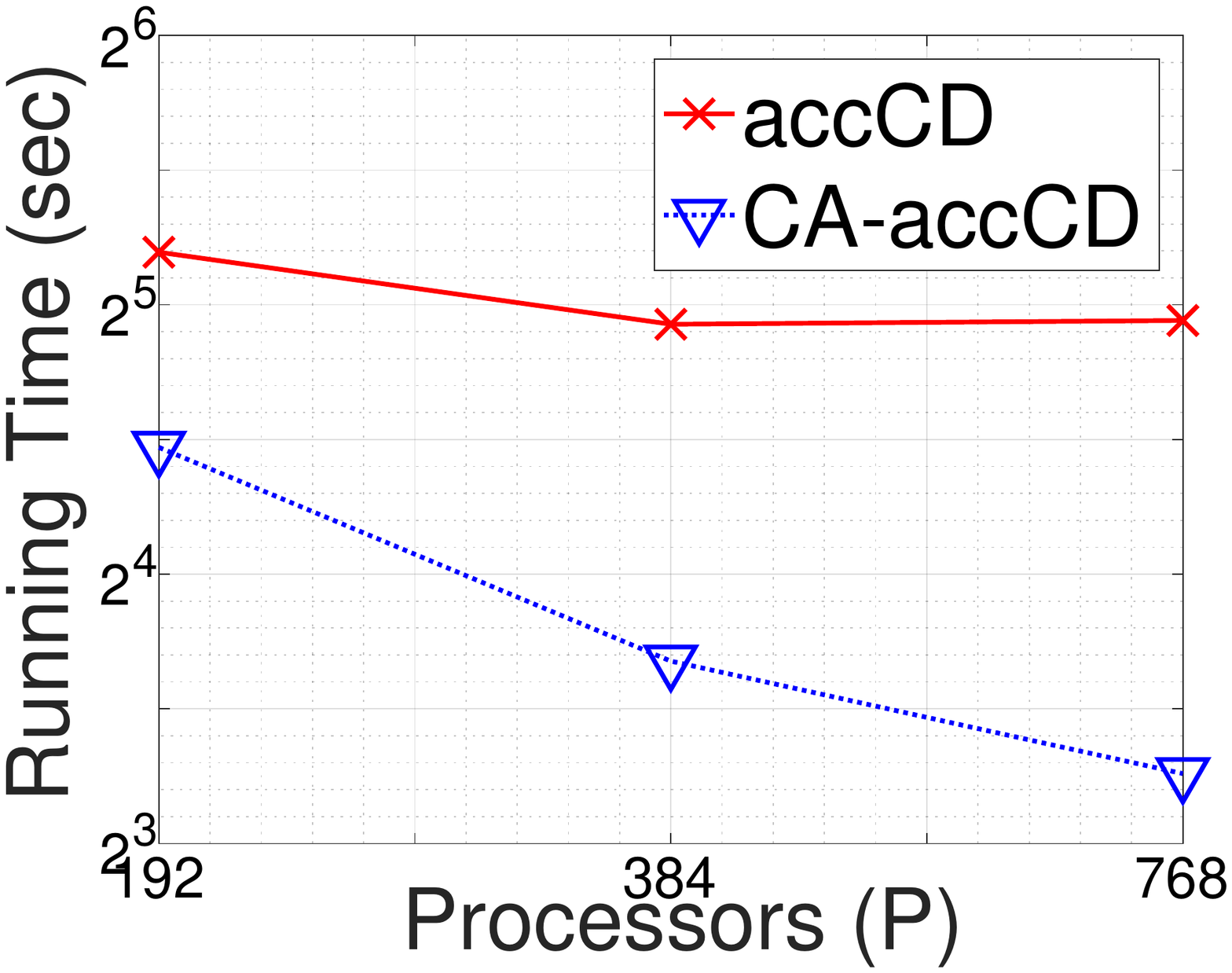}
\caption{news20 strong scaling}
\label{fig:news20scal}
\end{subfigure}
\begin{subfigure}{.24\textwidth}
\centering
\includegraphics[trim = 0.1in 2.5in 0.1in 2.5in, clip,width=1\textwidth]{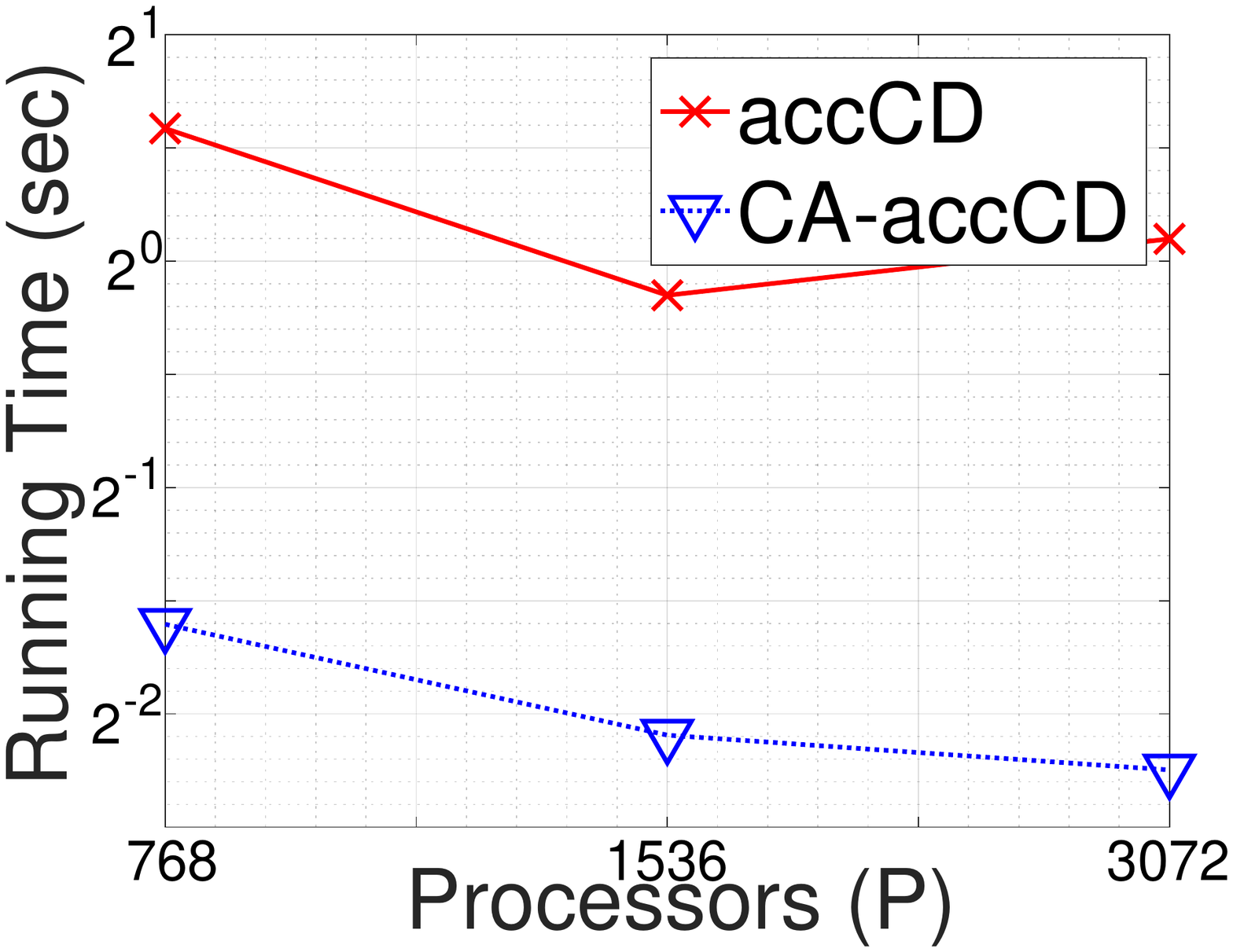}
\caption{covtype strong scaling}
\label{fig:covtypescal}
\end{subfigure}
\begin{subfigure}{.24\textwidth}
\centering
\includegraphics[trim = 0.1in 2.5in 0.1in 2.5in, clip,width=1\textwidth]{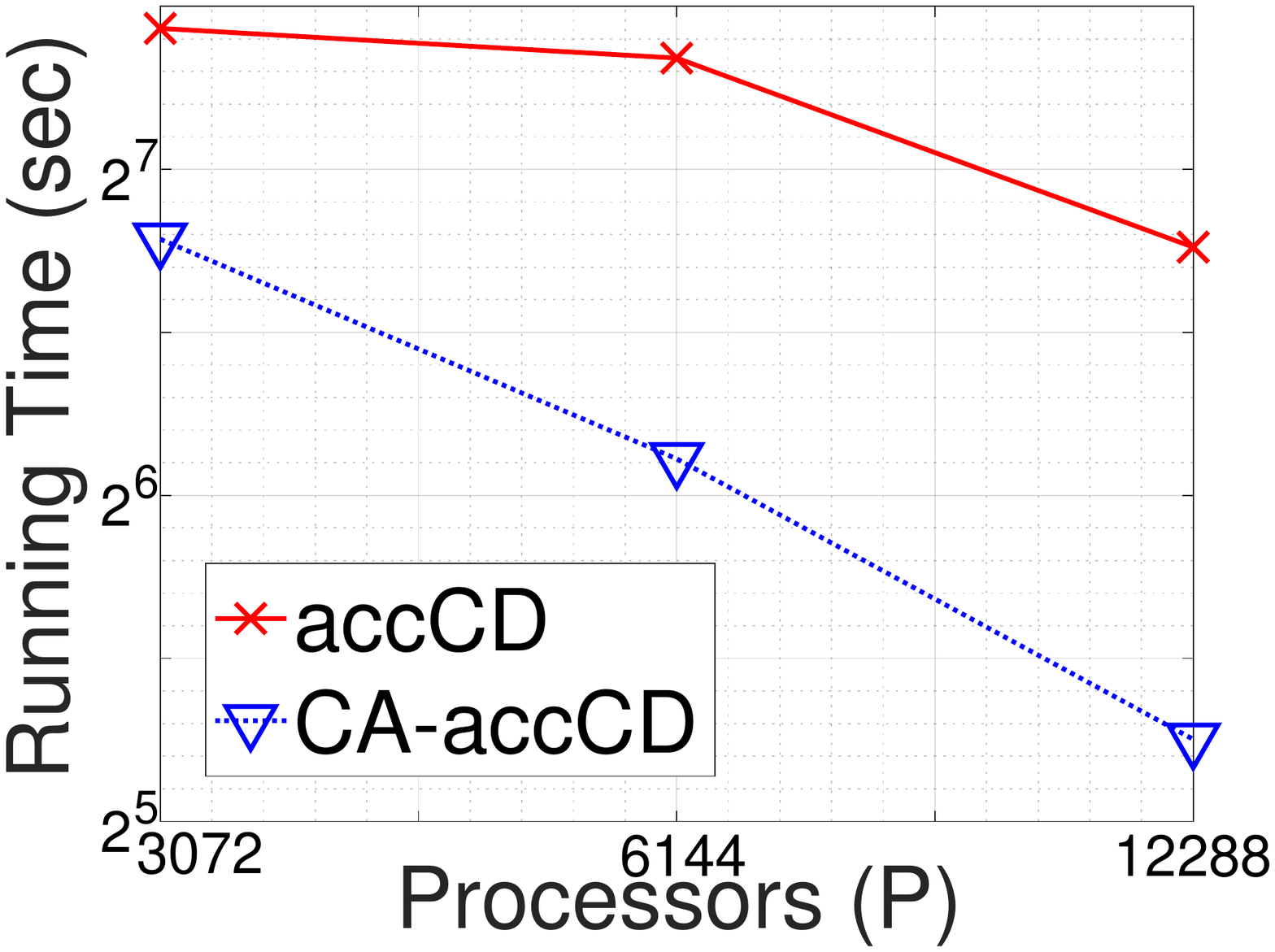}
\caption{url strong scaling}
\label{fig:urlscal}
\end{subfigure}
\begin{subfigure}{.24\textwidth}
\centering
\includegraphics[trim = 0.1in 2.5in 0.1in 2.5in, clip,width=1\textwidth]{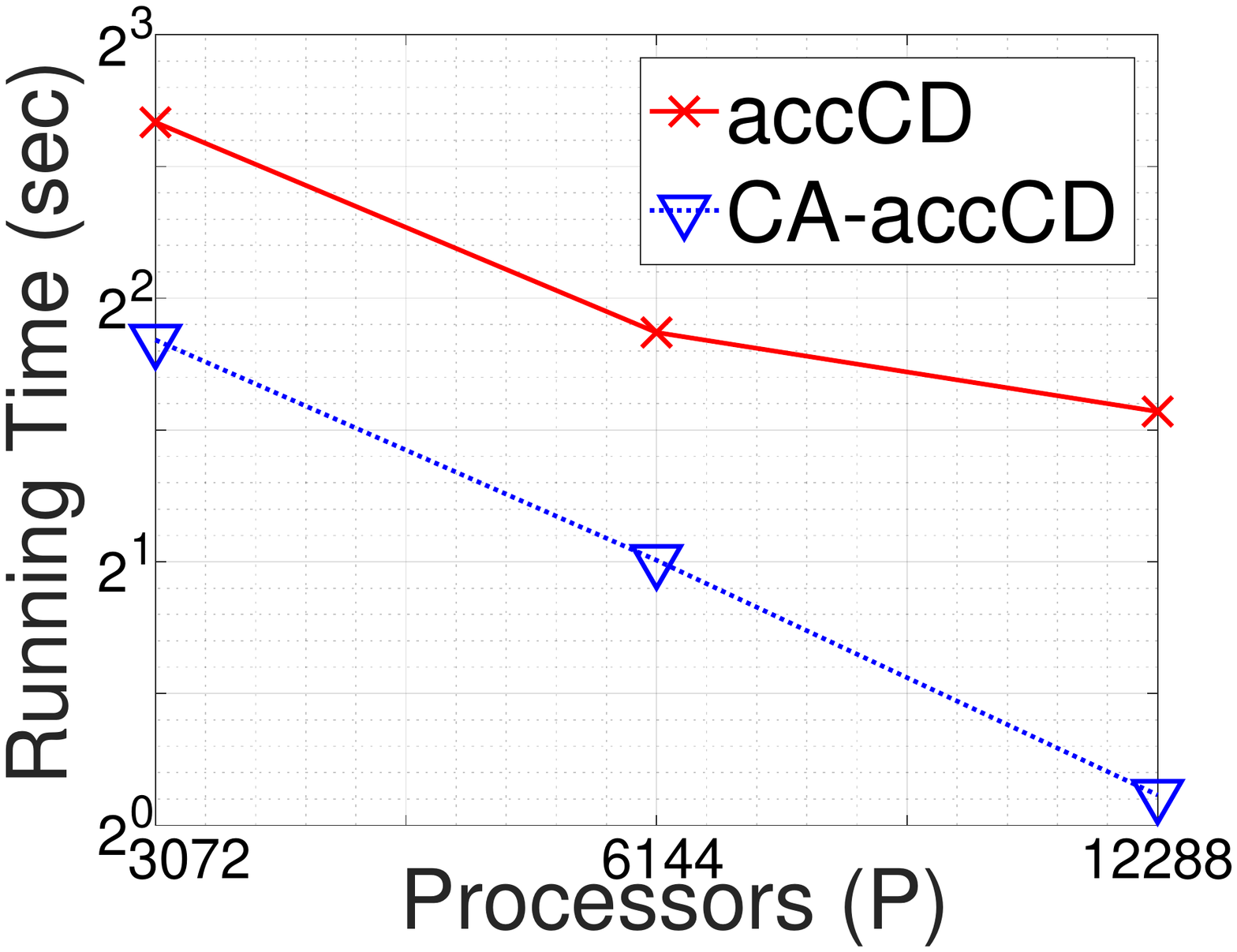}
\caption{epsilon strong scaling}
\label{fig:epsilonscal}
\end{subfigure}

\begin{subfigure}{.24\textwidth}
\centering
\includegraphics[trim = 0.1in 2.5in 0.1in 2.5in, clip,width=1\textwidth]{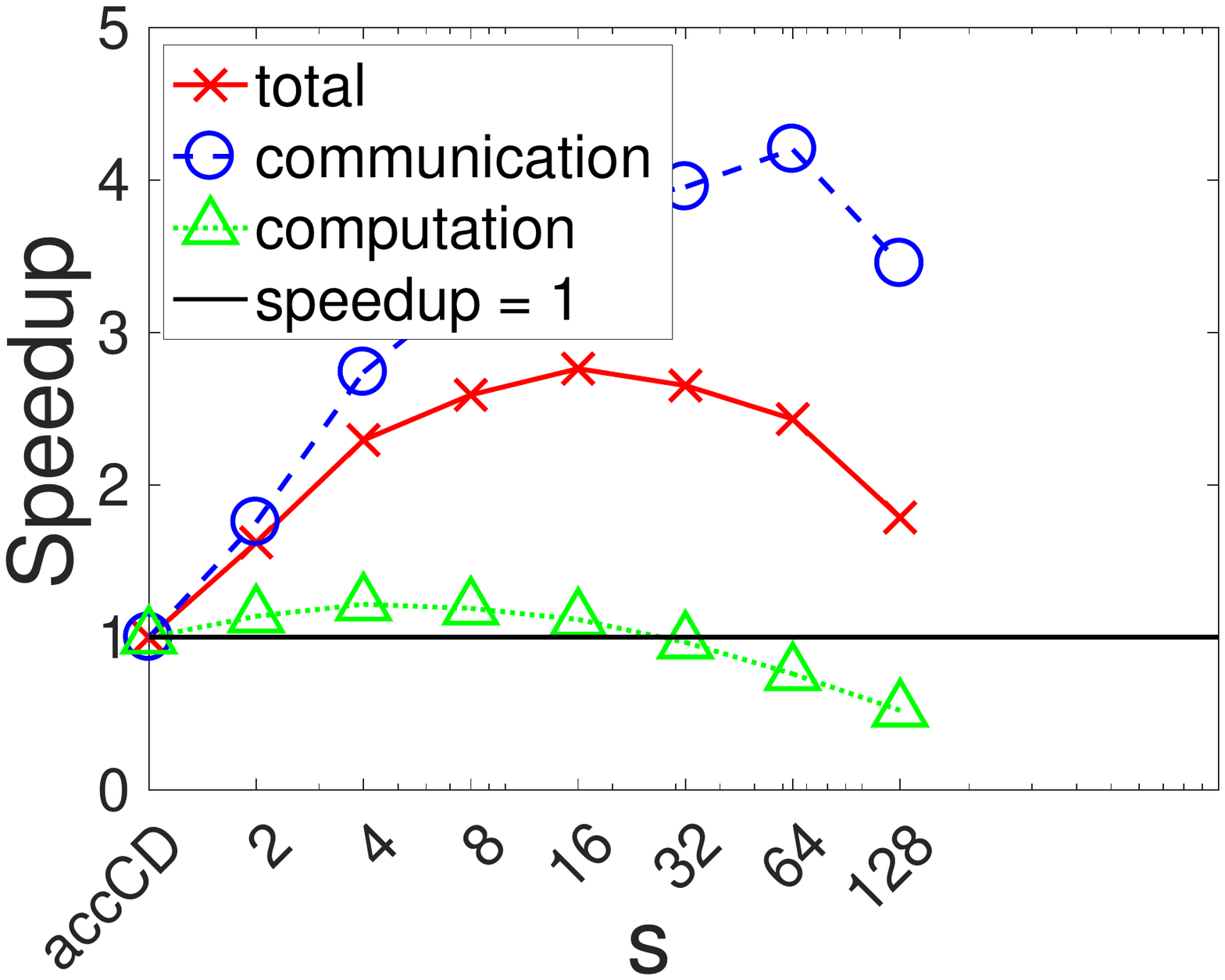}
\caption{news20 speedup}
\label{fig:news20sup}
\end{subfigure}
\begin{subfigure}{.24\textwidth}
\centering
\includegraphics[trim = 0.1in 2.5in 0.1in 2.5in, clip,width=1\textwidth]{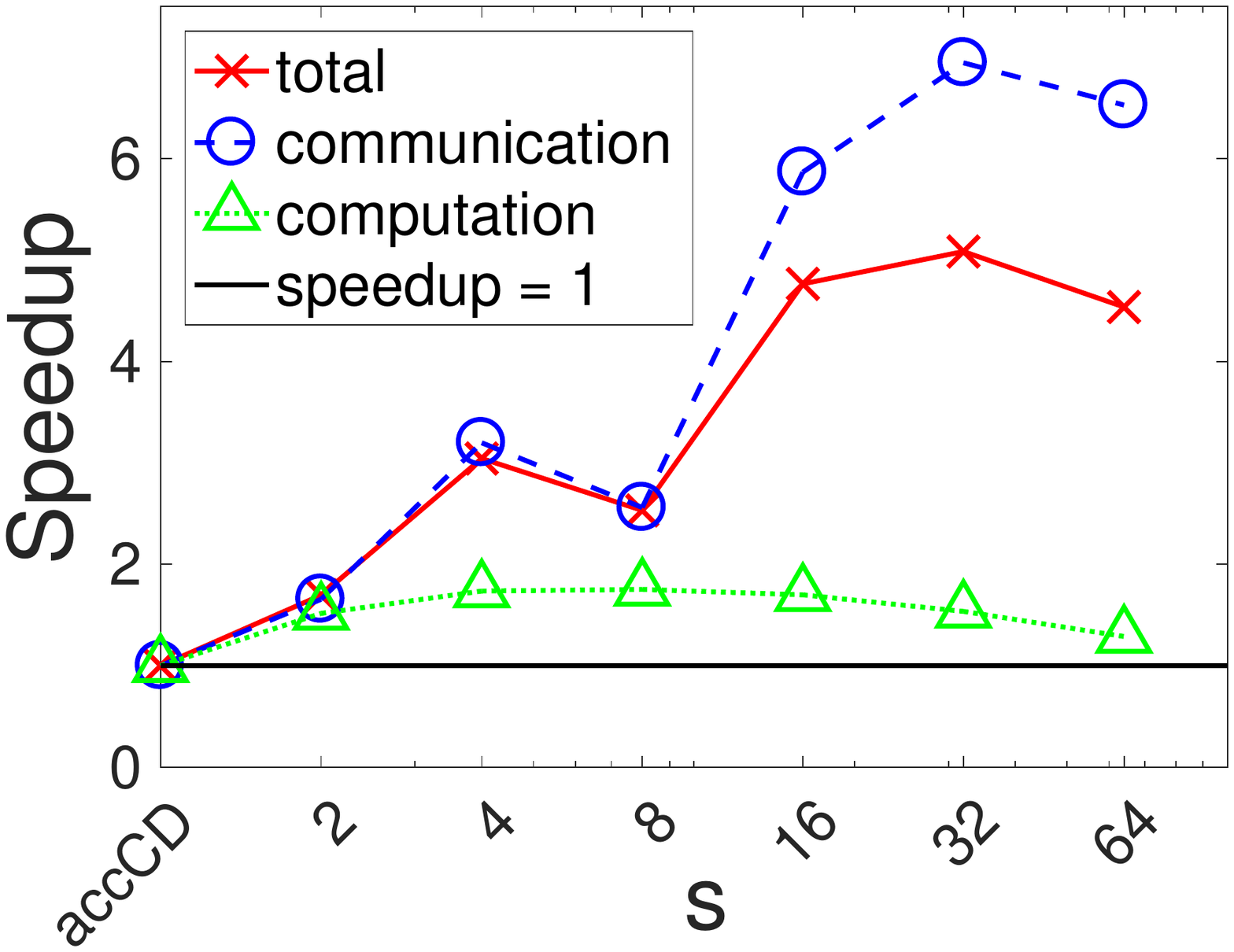}
\caption{covtype speedup}
\label{fig:covtypesup}
\end{subfigure}
\begin{subfigure}{.24\textwidth}
\centering
\includegraphics[trim = 0.1in 2.5in 0.1in 2.5in, clip,width=1\textwidth]{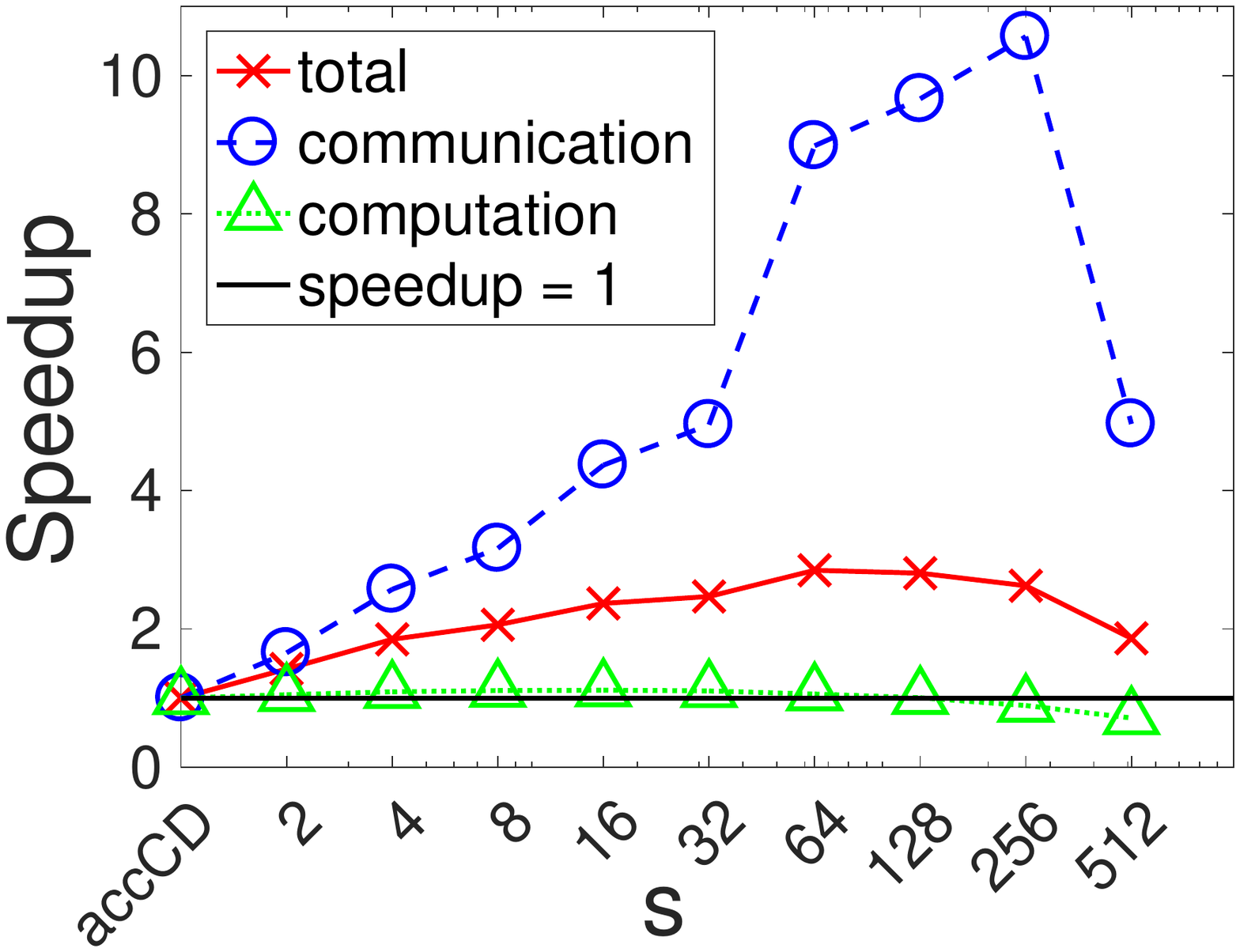}
\caption{url speedup}
\label{fig:urlsup}
\end{subfigure}
\begin{subfigure}{.24\textwidth}
\centering
\includegraphics[trim = 0.1in 2.5in 0.1in 2.5in, clip,width=1\textwidth]{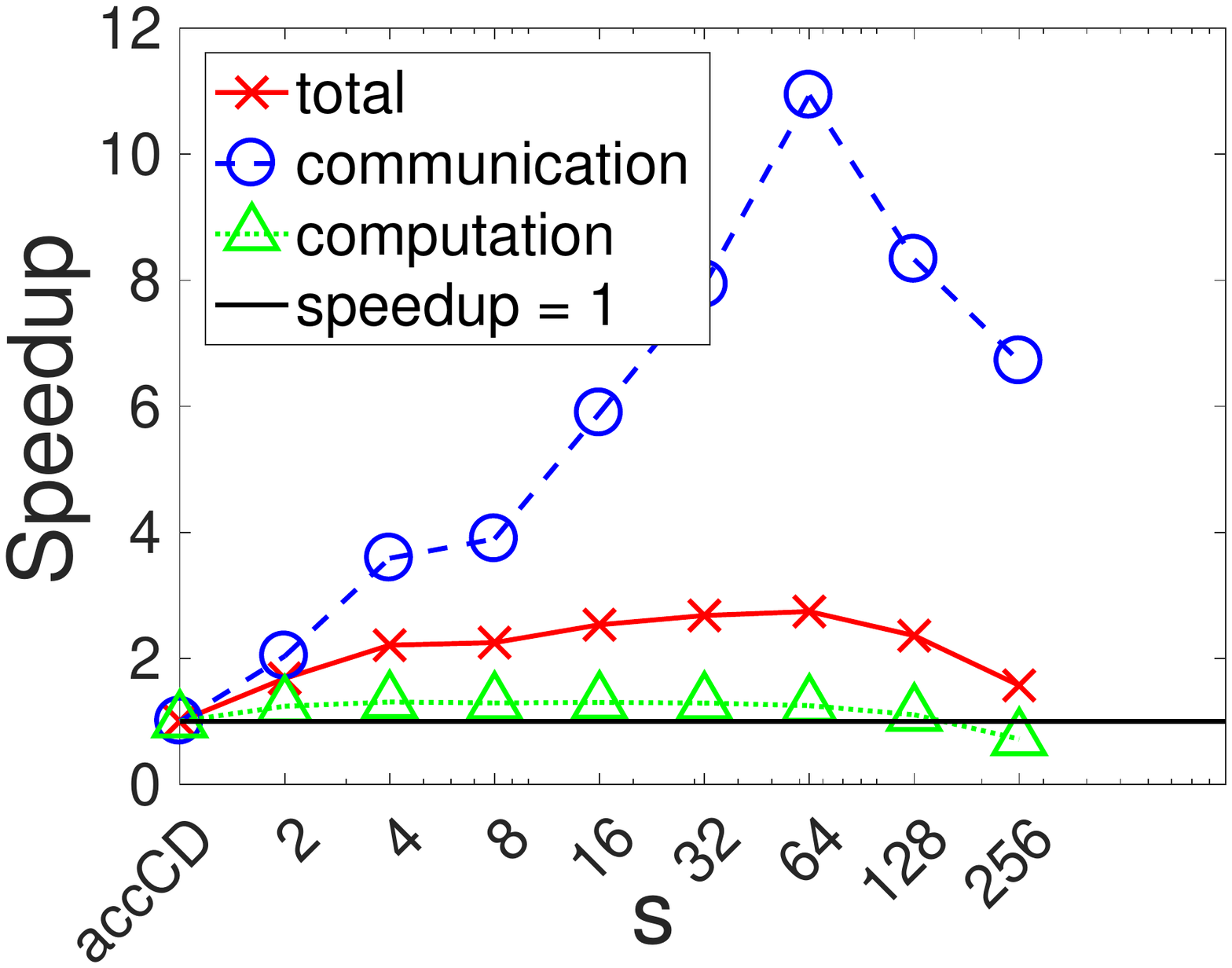}
\caption{epsilon speedup}
\label{fig:epsilonsup}
\end{subfigure}
\caption{We compare the strong scaling and speedups of SA-accCD in comparison to accCD on datasets in Table \ref{tbl:dsets}.}
\label{fig:scmp}
\end{figure*}

\paragraph{Convergence vs. Running Time}\label{sec:rtexp}
Figure \ref{fig:sobj} shows the convergence vs. running time for the datasets in Table \ref{tbl:dsets}. We present experiments on CD, accCD, BCD, accBCD and their SA variants. In all plots, we can observe that the accelerated methods converge faster than the non-accelerated methods. The BCD methods converge faster than the CD methods as expected. Since the SA-methods do not alter the convergence rates they are faster per iteration.

For the SA methods, we plot two values of $s$, one value (in blue) where we observed the best speedups and a larger value (in red) where we observed less speedups. Note that this decrease in speedup for certain values of $s$ is expected since the SA-methods tradeoff additional message size and computation for a decrease in latency cost. We see SA-accCD speedups of $2.8\times, 5.1\times, 2.8\times,$ and $2.7\times$ for news20, covtype, url, and epsilon, respectively. For SA-accBCD the speedups decrease to $1.2\times, 4.4\times, 2.1\times,$ and $1.8\times$, respectively. 

\paragraph{Performance Scaling and Speedups}\label{sec:supfexp}
Figure \ref{fig:scmp} shows the performance \emph{strong scaling} (problem size is fixed and number of processors is increased) and the breakdown of speedups. Figures \ref{fig:news20scal}-\ref{fig:epsilonscal} show the strong scaling performance of the accCD vs. SA-accCD methods for different ranges of processors for the datasets tested. We can observe that SA-accCD is faster for all datasets and for all processor ranges. Notice that the gap between accCD and SA-accCD (depicted in $\log_2$-scale) increases with the number of processors.

Figures \ref{fig:news20sup}-\ref{fig:epsilonsup} show the communication, computation, and total speedups attained by SA-accCD for several values of $s$ over accCD. We see large communication speedups which eventually decrease when message size costs dominate latency costs. SA-accCD also attains modest computation speedups over accCD. This is because selecting $s$ columns of $A$ and computing the $s^2$ entries of the Gram matrix (for SA-accCD) is more cache-efficient (uses a BLAS-3 routine) than computing $s$ individual dot-products (uses a BLAS-1 routine). Once $s$ becomes too large we see slowdowns.\label{sec:exp}
\section{Synchronization-Avoiding SVM}
We are given a matrix $A \in \mathbb{R}^{m \times n}$, labels $b \in \mathbb{R}^m$ where $b_i$ are binary labels $\{-1, +1\}$ for each observation $A_i$ ($i$-th row of $A$). Support Vector Machines (SVM) solve the optimization problem:
\begin{align}
\argmin_{x \in \mathbb{R}^n} \frac{1}{2}||x||_2^2 + \lambda \sum_{i =1}^m F(A_i,b_i,x)\label{eq:svmprimal}
\end{align}
wher $F(A_i,b_i,x)$ is a loss function and $\lambda > 0$ is the penalty parameter. In this work, we consider the two loss functions:
\begin{align}
\max(1 - b_iA_ix, 0) \quad \text{and} \quad \max(1 - b_iA_ix, 0)^2.\label{eq:svmdual}
\end{align}

We refer to the first as SVM-L1 and the second as SVM-L2 (consistent with \cite{dcdsvm08}). Recent work \cite{dcdsvm08} has shown that both variants of SVM can be solved efficiently using dual coordinate descent. Therefore, in this work we will consider the dual optimization problem:
\begin{align}
&\argmin_{\alpha \in \mathbb{R}^m} \frac{1}{2}\alpha^T \bar Q\alpha - e^T\alpha\\
&\text{subject to} \quad 0 \leq \alpha_i \leq \nu, \forall i,
\end{align}
\begin{algorithm}[t!]
  \caption{Dual Coordinate Descent for Linear SVM (SVM)}\label{alg:svm}
  \begin{algorithmic}[1]
    \State \textbf{Input}: $A \in \mathbb{R}^{m \times n}, b \in \mathbb{R}^m$, $H>1$, $\lambda \in \mathbb{R}$, $\alpha_0 \in \mathbb{R}^m$
    \State $x_0 = \sum_{j = 1}^m y_i\alpha_iA_i^T$
    \For {$h = 1\ldots H$}
      \State $i_h \in [m]$, chosen uniformly at random.
	  \State $\mathbb{I}_h = \left[e_{i_h}\right]$
	  \State Let $\mathbb{A}_{h} = \mathbb{I}_h^TA$
      \Statex  {\bf Communication: Lines 7 and 8.}
	  \State $\eta_h = \mathbb{A}_h\mathbb{A}_h^T + \gamma$
	  \State $g_h = \mathbb{I}_h^Tb\mathbb{A}_hx_{h-1}  - 1 + \gamma \mathbb{I}_h^T\alpha_{h-1}$\label{eq:gradrec}
	  \State $\tilde g_h = |\min(\max(\mathbb{I}_h^T\alpha_{h-1} - g_h, 0), \nu) - \mathbb{I}_h^T\alpha_{h-1}|$
      \If {$\tilde g_h \neq 0$}
		\State $\theta_h = \min(\max(\mathbb{I}_h^T\alpha_{h-1} - \frac{g_h}{\eta_{h}}, 0), \nu) - \mathbb{I}_h^T\alpha_{h-1}$
      \Else
        \State $\theta_h = 0$
      \EndIf
	  \State $\alpha_{h} =  \alpha_{h-1} + \theta_h \mathbb{I}_h$\label{eq:alprec}
	  \State $x_h = x_{h-1} + \theta_h \mathbb{I}_h^Tb \mathbb{A}_h^T$\label{eq:xrec}

    \EndFor
    \State \textbf{Output:} $x_H$
  \end{algorithmic}
\end{algorithm}
where $\bar Q = Q + D$, where $D = \gamma I_m$ and $Q_{ij} = b_ib_j A_iA_j^T$. For SVM-L1, $\gamma = 0$ and $\nu = \lambda$ and for SVM-L2 $\gamma = \frac{.5}{\lambda}$ and $\nu = \infty$. The dual problem can be solved efficiently by CD \cite{dcdsvm08} and is shown in Algorithm \ref{alg:svm}. Note that, unlike Lasso, SVM requires 1D-column partitioning in order to compute dot-products in parallel.
The recurrences defined in lines 7-11, 13, and 14 of Alg. \ref{alg:svm} can be unrolled to avoid synchornization. We begin the SA derivation by changing the loop index from $h$ to $sk + j$ where $k$ is the outer loop index, $s$ is the (tunable) recurrence unrolling parameter, and $j$ is the inner loop index. Let us assume that we are at iteration $sk + 1$ and have just computed the vectors $x_{sk}$ and $\alpha_{sk}$. From this the next update, $\theta_{sk + 1}$, can be computed by

\begin{align*}
	g_{sk + 1} &= \mathbb{I}_{sk + 1}^Tb\mathbb{A}_{sk + 1}x_{sk}  - 1 + \gamma \mathbb{I}_{sk+1}^T\alpha_{sk},\\
	\tilde g_{sk + 1}& = |\min(\max(\mathbb{I}_{sk+1}^T\alpha_{sk} - g_{sk+1}, 0), \nu) - \mathbb{I}_{sk+1}^T\alpha_{sk}|,\\
\theta_{sk + 1} &=
\left\{
        \begin{array}{ll}
			\min(\max(\mathbb{I}_{sk+1}^T\alpha_{sk} - \frac{g_{sk + 1}}{\eta_{sk+1}}, 0), \nu) - \mathbb{I}_{sk+1}^T\alpha_{sk},\\ \quad\quad \quad\quad \quad\quad\text{when} ~ \tilde g_{sk + 1} \neq 0\\
            0,\quad\quad\quad\quad\quad \text{otherwise}.
        \end{array}
    \right.
\end{align*}
Finally, $\alpha_{sk +1}$ and $x_{sk+1}$ can be computed by
\begin{align*}
	\alpha_{sk + 1} &=  \alpha_{sk} + \theta_{sk + 1} \mathbb{I}_{sk+1},\\
	x_{sk + 1} &= x_{sk} + \theta_{sk + 1} \mathbb{I}_{sk+1}^Tb \mathbb{A}_{sk + 1}^T.
\end{align*}

\begin{algorithm}[t!]
  \caption{Synchronization-Avoiding Linear SVM (SA-SVM)}\label{alg:casvm}
  \begin{algorithmic}[1]
    \State \textbf{Input}: $A \in \mathbb{R}^{m \times n}, b \in \mathbb{R}^m$, $H>1$, $\lambda \in \mathbb{R}$, $s \in \mathbb{Z}^+$ $\alpha_0 \in \mathbb{R}^m$,
    \Statex $\gamma =  \left\{ \begin{array}{ll}0, \text{for SVM-L1}\\ \frac{.5}{\lambda}, \text{for SVM-L2}\end{array}\right.$, $\nu =  \left\{ \begin{array}{ll}\lambda, \text{for SVM-L1}\\ \infty, \text{for SVM-L2}\end{array}\right. \vspace{1mm} $
    \State$x_0 = \sum_{i = 1}^m y_i\alpha_iA_i^T$
    \For {$k = 0,\ldots, \frac{H}{s}$}
      \For {$j = 1\ldots s$}
        \State $i_{sk + j} \in [m]$, chosen uniformly at random.
		\State Let $\mathbb{I}_{sk + j} = \left[e_{i_{sk + j}}\right]$
		\State Let $\mathbb{A}_{{sk + j}} = \mathbb{I}_{sk + j}^T A$
        \EndFor
	  \State Let $Y = \quad \left[\mathbb{A}_{sk+1}^T, \mathbb{A}_{sk + 2}^T, \ldots, \mathbb{A}_{sk + j}^T\right]$.
      \Statex {\bf Communication: Lines 9 and 10.}
      \State $G = Y^TY + \gamma I_s$.
      \State $[x'_{sk + 1},\ldots, x'_{sk + s}]^T = Y^Tx_{sk}$
      \State $[\eta_{sk +1},\ldots, \eta_{sk+s}]^T = \diag(G)$
      \For {$j = 1,\ldots, s$}
        \State Compute $\beta_{sk + j}$ according to equation \eqref{eq:sbetrec}.
        \State Compute $g_{sk + j}$ according to equation\footnotemark \eqref{eq:sgradrec}.
        \State $\tilde g_h = |\min(\max(\beta_{sk + j} - g_{sk + j}, 0), \nu) - \beta_{sk+j}|$
        \If {$\tilde g_h \neq 0$}
          \State $\theta_{sk + j} = \min(\max(\beta_{sk + j} - \frac{g_{sk + j}}{\eta_{sk + j}}, 0), \nu)$
          \Statex$\quad \quad \quad \quad \quad \quad \quad - \beta_{sk + j}$
        \Else
          \State $\theta_{sk + j} = 0$
        \EndIf
	  \State $\alpha_{sk + j} =  \alpha_{sk + j - 1} + \theta_{sk + j} \mathbb{I}_{sk + j}$
	  \State $x_{sk + j} = x_{sk + j -1} + \theta_{sk + j} \mathbb{I}_{sk + j}^Tb \mathbb{A}_{sk+j}^T$
      \EndFor
    \EndFor
    \State \textbf{Output:} $x_H$
  \end{algorithmic}
\end{algorithm}
By unrolling the vector update recurrences for $\alpha_{sk +1}$ and $x_{sk+1}$, we can compute $g_{sk + 2}$, $\tilde g_{sk + 2}$, and $\theta_{sk + 2}$ in terms of $\alpha_{sk}$ and $x_{sk}$. We will ignore the quantities $\tilde g_{sk + j}$ and $\theta_{sk+j}$ in the subsequent derivations for brevity. We introduce an auxiliary scalar, $\beta_{sk + j} = \mathbb{I}_{sk + j}^T\alpha_{sk + j - 1}$, for notational convenience.
\begin{align*}
	\beta_{sk + 2} &= \mathbb{I}_{sk + 2}^T\alpha_{sk} + \theta_{sk + 1}\mathbb{I}_{sk + 2}^T\mathbb{I}_{sk + 1},\\
\begin{split}
	g_{sk + 2} &= \mathbb{I}_{sk + 2}^Tb{\mathbb{A}_{sk + 2}}x_{sk}  - 1 + \gamma \beta_{sk + 2}\\ 
&+\theta_{sk + 1}\mathbb{I}_{sk + 2}^Tb\mathbb{I}_{sk + 1}^Tb{\mathbb{A}_{sk + 2}}\mathbb{A}_{{sk + 1}}^T .\end{split}
\end{align*}

By induction we can show that $g_{sk + j}$ can be computed in terms of $\alpha_{sk}$ and $x_{sk}$ such that
\begin{align}
	\beta_{sk + j}  &= \mathbb{I}_{sk + j}^T\alpha_{sk} + \sum_{t = 1}^{j-1} \theta_{sk + t}\mathbb{I}_{sk + j}^T\mathbb{I}_{sk + t} ,\label{eq:sbetrec}\\
\begin{split}
	g_{sk + j} &= \mathbb{I}_{sk + j}^Tb\mathbb{A}_{sk + j}x_{sk}  - 1 + \gamma \beta_{sk +j} \\
	&+ \sum_{t = 1}^{j -1} \theta_{sk + t}\mathbb{I}_{sk + j}^Tb \mathbb{I}_{sk + t}^Tb \mathbb{A}_{sk + j}\mathbb{A}_{sk + t}^T,\label{eq:sgradrec}
\end{split}
\end{align}
\footnotetext{Since $\mathbb{A}_{sk+j}x_{sk+j} = x'_{sk + j}$, no additional computation or communication is needed to form the vector.}%
for $j = 1,2,\ldots, s$. 
Due to the recurrence unrolling, we can defer updates to $\alpha_{sk}$ and $x_{sk}$ for $s$ iterations. The summation in \eqref{eq:sbetrec} adds a previous update $\theta_{sk + t}$ if the coordinate chosen for update at iteration $sk + t$ is the same as iteration $sk + j$. Synchronization can be avoided in this step by initializing the random number generator on all processors to the same seed. The summation in \eqref{eq:sgradrec} computes the inner-product $\mathbb{A}_{sk + j}\mathbb{A}_{sk + t}^T$. Synchronization can be avoided at this step by computing the Gram-like matrix, $\left[\mathbb{A}_{sk + 1},\ldots, \mathbb{A}_{sk + s}\right]\left[\mathbb{A}_{sk + 1}^T,\ldots, \mathbb{A}_{sk + s}^T\right]$, upfront at the beginning of the outer loop. Note that the diagonal elements of the resulting matrix are the $\eta_{sk + j}$'s required in the inner loop. Finally, at the end of the $s$ inner loop iterations we can perform the vector updates: $\alpha_{sk + s} =  \alpha_{sk} + \sum_{t = 1}^s\theta_{sk + t} \mathbb{I}_{sk + t}$ and $x_{sk + s} = x_{sk} + \sum_{t = 1}^s\theta_{sk + t} \mathbb{I}_{sk + t}^Tb \mathbb{A}_{sk + t}^T$.

The resulting SA-SVM algorithm is shown in Alg. \ref{alg:casvm}. The derivation we present in this section only rearranges the algebra. Hence, the convergence rates and behavior of SVM (Alg. \ref{alg:svm}) do not change (in exact arithmetic). However, in floating-point arithmetic this rearrangement could lead to numerical instability. However, we will empirically show in Section \ref{sec:exp} that SA-SVM is numerically stable. While our SA-variants reduce the latency cost by $s$, they increase the flops and bandwidth costs by $s$ (due to the $s \times s$ Gram matrix, $G$). Therefore, the best choice of $s$ depends on the relative algorithmic flops, bandwidth, latency costs and their respective hardware parameters.

\label{sec:deriv_svm}
\section{Experimental Results: SVM}
\begin{table}[t]
  \centering
\begin{tabular}{l|c|c|c}
\hline
\multicolumn{4}{c}{Summary of datasets}\\
\hline\hline
\multicolumn{1}{c}{Name} &  \multicolumn{1}{|c|}{Features} &Data Points & \multicolumn{1}{|c}{NNZ$\%$}\\
\hline
w1a & $2,477$ & $300$ & $4$\\ \hline
leu &$7,129$ & $38$ & $100$\\ \hline
duke & $7,129$ & $44$ & $100$\\ \hline
news20.binary & $19,996$ & $1,355,191$ & $0.03$\\ \hline
rcv1.binary & $20,242$ & $47,236$ & $0.16$\\ \hline
gisette & $6000$ & $5000$ & $99$\\ \hline
\end{tabular}
\caption{Properties of the LIBSVM datasets used.}
\label{tbl:svmdsets}
\end{table}
The recurrence unrolling results in the computation of an $s \times s$ Gram matrix, whose condition number may adversely affect numerical stability. So, we begin by verifying the stability of SA-SVM through MATLAB experiments and then illustrate the speedups attainable through our approach.

\begin{figure}[t!]
\centering
\begin{subfigure}{.24\textwidth}
\centering
\includegraphics[trim = 0.45in 2.5in 0.4in 2.5in, clip,width=1\textwidth]{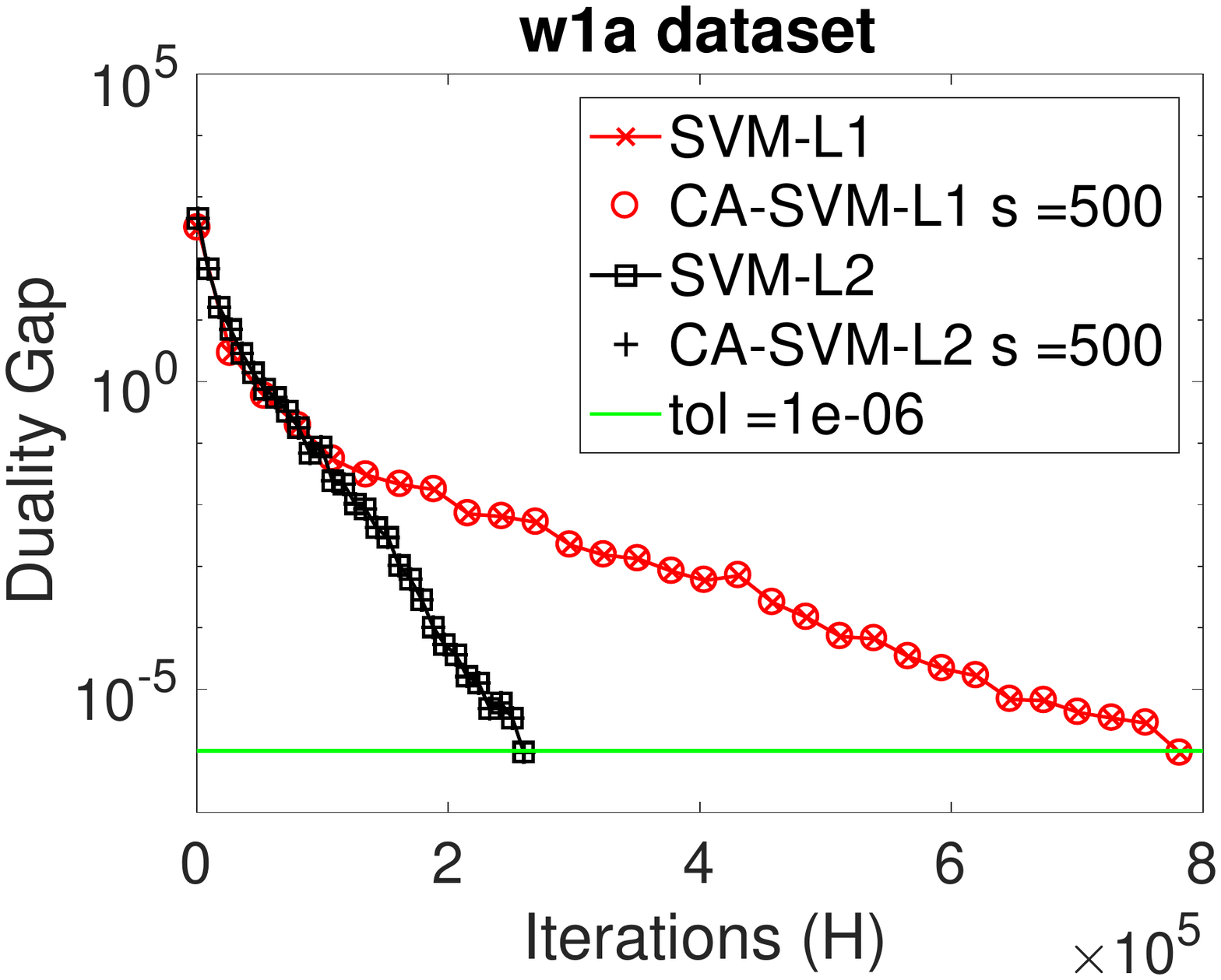}
\label{fig:w1a}
\end{subfigure}%
\begin{subfigure}{.24\textwidth}
\centering
\includegraphics[trim = 0.45in 2.5in 0.4in 2.5in, clip,width=1\textwidth]{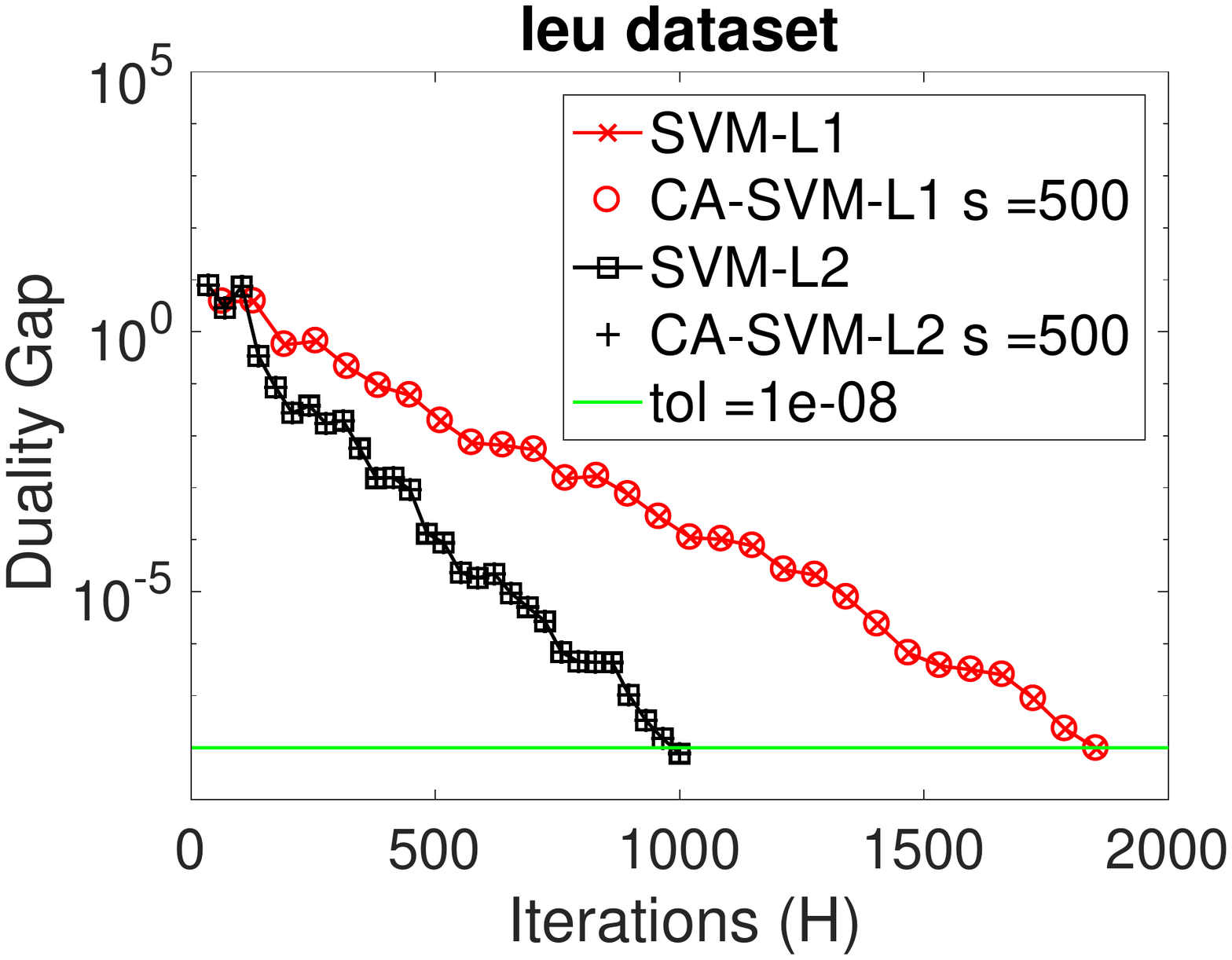}
\label{fig:leu}
\end{subfigure}

\begin{subfigure}{.24\textwidth}
\centering
\includegraphics[trim = 0.45in 2.5in 0.4in 2.5in, clip,width=1\textwidth]{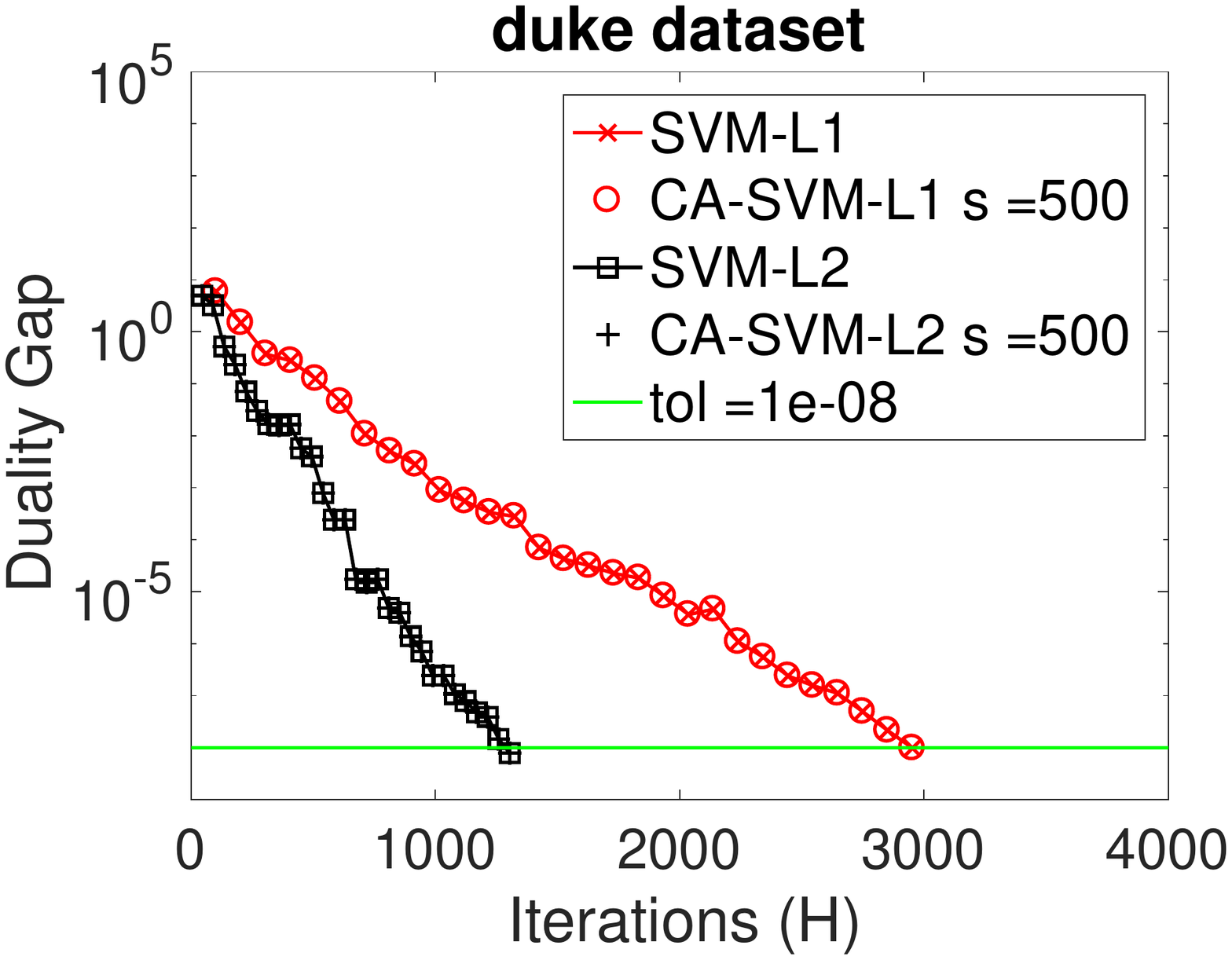}
\label{fig:duke}
\end{subfigure}
\caption{Duality gap vs. iterations of SVM-L1, SVM-L2, and their SA variants with $s = 500$.}
\label{fig:svmconv}
\end{figure}

We conduct numerical stability experiments in MATLAB (same platform as Sec. \ref{sec:exp}) and use binary classification datasets (summarized in Table \ref{tbl:svmdsets}) from the LIBSVM \cite{cc01} repository. We measure the convergence behavior by plotting the duality gap, $P(x) - D(\alpha)$, where $P(x)$ is the primal objective value\footnote{We can compute this without running the primal SVM algorithm (as in \cite{dcdsvm08}).} and $D(\alpha)$ is the dual objective value (as in \cite{dcdsvm08}). Note that duality gap is a stronger criterion than the relative objective error used in Sec. \ref{sec:exp}. Due to strong convexity, primal and dual linear SVM have the same optimal function value \cite{dcdsvm08,platt98}. 
We set $\lambda = 1$ for all experiments (same as \cite{dcdsvm08}) and show results for SVM-L1 and SVM-L2.
\begin{table}
\begin{center}
\begin{tabular}{c|c|c|c}
\hline
\multicolumn{4}{c}{SA-SVM Speedups}\\
\hline\hline
Dataset & Processors & Algorithm & Running Time (speedup) \\
\hline
\multirow{2}{*}{news20.binary} & \multirow{2}{*}{P = 576} & SVM-L1 & $258$  sec.\\ \cline{3-4}
& & SA-SVM-L1& \bm{$121$}  {\bf sec. (\bm{$2.1\times$})}\\ \hline
\multirow{2}{*}{rcv1.binary} & \multirow{2}{*}{P = 240} & SVM-L1 & $208$  sec.\\ \cline{3-4}
& & SA-SVM-L1& \bm{$149$}  {\bf sec. (\bm{$1.4\times$})}\\ \hline
\multirow{2}{*}{gisette} & \multirow{2}{*}{P = 3072} & SVM-L1 & $230$  sec.\\ \cline{3-4}
& & SA-SVM-L1& \bm{$57$}  {\bf sec. (\bm{$4\times$})}\\ \hline
\end{tabular}
\end{center}
\caption{SA-SVM-L1 speedups over SVM-L1. $s = 64$ was the best setting for rcv1 and news20 datasets; $s = 128$ was best for gisette. We use a duality gap tolerance of $1e{-}1$.}
\label{tbl:svmsup}
\end{table}

Figure \ref{fig:svmconv} illustrates that the SA-variants are numerically stable and converge in the same way their non-SA variants. SVM-L2 converges faster than SVM-L1 since the loss function is smoothed. Table \ref{tbl:svmsup} shows speedups of SA-SVM-L1 over SVM-L1. SVM-L1 and SVM-L2 are initialized with different scalar quantities. All else remains the same, so we solve the (harder) SVM-L1 problem and report performance results. Note that we perform offline strong scaling experiments for each dataset and report the best processors and running time combinations. We observed speedups of $1.4\times$ for rcv1.binary, $2.1\times$ for news20.binary and $4\times$ for gisette. These speedups were attained despite load balancing issues for rcv1 and news20 when transforming datasets stored row-wise on disk to 1D-column partitioned matrices in DRAM (Lasso experiments do not have these issues). Eliminating this overhead in future work would further improve speedups and scalability (since load imbalance decreases the effective flops rate due to stragglers). Gisette is nearly dense hence load balance was not a problem. We were able to strong scale SA-SVM-L1 to $3072$ cores and attain a $4\times$ speedup.
\label{sec:exp_svm}
\section{Conclusion}
We showed in this paper that existing work in CA-NLA can be extended to important ML problems. We derived SA variants of CD and BCD methods and illustrated that our methods are faster and more scalable with changing convergence rates. The SA-variants attain speedups of $1.2\times$ - $5.1\times$ and can reduce communication by factors of $4.2\times$ - $10.9\times$. While we did not explore other parallel environments, our methods would attain greater speedups on frameworks like Spark \cite{spark} due to the large latency costs \cite{gittens16}.\label{sec:conc}

\bibliographystyle{IEEEtran}
\bibliography{refs}
%

\end{document}